\documentclass[11pt,submission]{IEEEtran}

\newtheorem{theorem}{Theorem}
\title{Classes of lower bounds on outage error probability and {MSE} in {B}ayesian parameter estimation}

\author{Tirza Routtenberg and Joseph Tabrikian \\ Department of Electrical and Computer Engineering,  \\ Ben-Gurion University of the Negev
 Beer-Sheva 84105, Israel \\
 Phone: $+$\,972\,-86477774,
 Email: \{tirzar,joseph\}@ee.bgu.ac.il}

\usepackage{amssymb, amsmath}
\usepackage{graphicx}
 \usepackage[dvips]{color}
 \newcommand{\ud}{\,\mathrm{d}}

\newcommand{\xvec}{{\bf{x}}}

\newcommand{\Fmat}{{\bf{F}}}

\newcommand{\Hmat}{{\bf{H}}}

\newcommand{\define}{\stackrel{\triangle}{=}}

\newcommand{\be}{\begin{equation}}
\newcommand{\ee}{\end{equation}}
\newcommand{\beqna}{\begin{eqnarray}}
\newcommand{\eeqna}{\end{eqnarray}}

\input{psfig.sty}
\begin{document}
\maketitle
\nopagebreak

\begin{abstract}
In this paper, new classes of lower bounds on the outage error probability  and  on  the  mean-square-error (MSE) in Bayesian parameter estimation  are
proposed.  The minima of the $h$-outage error probability and the MSE are obtained by
the  generalized maximum
{\em a-posteriori} probability  and the minimum MSE (MMSE) estimators, respectively.
However, computation of these estimators and  their corresponding performance is usually not tractable and thus, lower bounds on these terms can be very useful for performance analysis.
The proposed class of lower bounds on the outage error probability     is derived using  H$\ddot{\text{o}}$lder's inequality.
This class  is utilized to derive a new class of  Bayesian MSE bounds.
It is shown that for unimodal symmetric conditional probability density functions (pdf)
 the tightest probability of outage error  lower bound in the proposed class attains the minimum probability of outage error and
the tightest MSE bound coincides with the MMSE performance. 
In addition, it is shown that the proposed MSE bounds are always tighter than  the Ziv-Zakai  lower
bound (ZZLB). The proposed bounds are compared with other existing performance lower bounds via some  examples.

\end{abstract}

\section{Introduction}
\label{sec:intro}
The  mean-square-error (MSE)  criterion has been commonly used  for performance analysis in parameter estimation.
Lower bounds on the MSE  are widely used for problems where the exact
minimum MSE (MMSE) is difficult to evaluate.  %These bounds can be used to investigate the properties of the optimal estimators (in the MMSE sense)  or  for assessing and comparing
%the performance of  estimators.
 Bayesian  MSE bounds  for random  parameters estimation  can be divided into two classes. The Weiss-Weinstein class
is based on the covariance inequality which includes the Bayesian CRB (BCRB) \cite{van}, the Reuven-Messer bound \cite{ReuvenMesser},  the Weiss-Weinstein lower
bound (WWLB), and the Bayesian Todros-Tabrikian bound \cite{TT}. The Ziv-Zakai class of bounds relates  the MSE in
the estimation problem to the probability of error in a binary
detection problem. The Ziv-Zakai class  includes the Ziv-Zakai lower
bound (ZZLB)  \cite{ZZLB1} and its improvements, notably the Bellini-Tartara bound \cite{BT74}, the Chazan-Zakai-Ziv bound \cite{ChazanZZ},
and Bell-Ziv-Zakai bound \cite{ExtendedZZ}.

Additional important criterion for performance analysis in parameter estimation is
the probability of outage error, which is the probability that the estimation error is higher than a given threshold. This criterion provides meaningful information in the presence of large errors, 
while the  occurrence of  large
errors with small probability may cause the MSE criterion to be non-informative. In some parameter estimation problems  we may be interested in  evaluation of outage error rate  and  the exact value of
the  error may be non-informative \cite{extractSig}, \cite{worst_case}.
The probability of outage error criterion
as a function of threshold error  provides information on the error distribution while the 
MSE  provides information only on the  second order moment of the estimation error. 
In addition, in many 
estimation problems the MSE is subject to a threshold  phenomenon 
 which determines   operation region  (see papers in \cite{van_bell}). 
 Thus,
 the  MSE threshold may be highly influenced by large errors with small probability 
 and thus the outage error probability criterion can be more useful for this propose.

%%%%%%%%%%%%%%%%%%%%%%%%%%%%%%%%%%%%%%%%%%%%%%%%%%%%%%%%%%%%%%%%%%%%%%%%%%%%%%%%%%  
The minimum outage error probability can be obtained by the generalized maximum
{\em a-posteriori} probability (MAP)  estimator which is given by maximization of the posterior function convolved with a rectangular window.
 However,
computation of the minimum outage error probability is usually not tractable, and
thus   tight lower bounds on the probability of outage error  are useful for performance
analysis and system design. In the literature,  only few  lower bounds on the probability of outage error can be found and most of them are based on the probability of error in binary or multiple hypothesis testing  problems.
 A known  lower bound for uniformly distributed unknown parameters is given by the Kotel'nikov's inequality \cite{Kotel}.
There are several works on   approximations of the probability of outliers for non-Bayesian direction-of-arrival (DOA)   estimation problem
(see e.g. \cite{AbramovichICASSP}, \cite{Athley}). In general,  the internal terms in the integral version of the ZZLB can be used  as lower bounds on the outage error probability.
In similar, MSE lower bounds can be obtained from outage error lower bounds by using the Chebyshev's inequality,  as in the original ZZLB \cite{ZZLB1},
or by using the  probability  identity  \cite{statistics}, as in \cite{BT74},\cite{ExtendedZZ}.
The Chebyshev's inequality is known to be unachievable and thus the second option is preferred.
The outage error probability can be interpreted as the probability of error for estimation problems. General classes of lower bounds for the probability of error in multiple hypothesis problems have been derived in \cite{my_detection}.

 %%%%%%%%%%%%%%%%%%%%%%%%%%%%%%%%%%%%%%%%%%%%%%%%%%%%%%%%%%%%
 
In this paper, a new class of Bayesian lower bounds on the minimum probability of outage error  is
derived using H$\ddot{\text{o}}$lder's inequality. 
In some cases, the proposed outage error probability bounds  are simpler to compute than the
minimum outage error probability   and they provide a good prediction of
this criterion.   It is shown that for parameter estimation problems with unimodal symmetric conditional probability density function (pdf), the tightest lower bound under this
class   coincides
with the optimum probability of outage error provided by the generalized  
MAP criterion.
In addition,   using the  probability  identity  \cite{statistics} new classes of Bayesian lower bounds on arbitrary  distortion measures
are derived. These  classes are based on  the minimum outage error probability bounds, derived in the first part of this paper, and on the probability identity  \cite{statistics}.  
For the new class of MSE lower bounds,
it is shown that the tightest bound in this class is always tighter than the ZZLB.   For parameter estimation problems with unimodal symmetric conditional pdf, the tightest bound in the proposed class   attains the MMSE.

%%%%%%%%%%%%%%%%%%%%%%%%%%%%%%%%%%%%%%%%%%%%%%%%%%%%%%%%%%%%%%%%%%%%%%%%%%%%%%%%
The paper is organized as follows. The  problem statement is presented in Section \ref{problem_S}. In Section
\ref{der}, 
the new class of Bayesian lower bounds on the probability of outage error   is derived and in Section \ref{tight_der}
the tightest subclass of lower bounds in this class is found. A new class of lower bounds on different distortion measures 
is derived in
Section \ref{MSE_bound} using the bounds on the probability of outage error. In particular, a
 new class of MSE Bayesian lower bounds is derived.
The bounds properties are described in Section \ref{proper}, and
the  performance of the proposed bounds  are evaluated in Section \ref{simulations} via some examples. Finally, our conclusions appear in Section \ref{diss}.

%{The tightest subclass of lower bounds on the probability of outage error}
%\label{tight_der}
%%%%%%%%%%%%%%%%%%%%%%%%%%%%%%%%%%%%%%%%%%%%%%%%%%%%%%%%%%%%%%%%%%%%%%%%%%%%%%%%%%%%%

\section{Problem statement}
\label{problem_S}
Consider the estimation of a continuous  scalar random variable $\theta\in{\mathbb{R}}$,
based on a random observation vector $\xvec\in\chi$ with the cumulative distribution function (cdf) $F_\xvec(\cdot)$ and  $F_{\theta|\xvec}(\cdot|\xvec)$ denotes the
conditional cdf of $\theta$ given $\xvec$. It is assumed that $F_{\theta|\xvec}(\cdot|\xvec)$ is continuous such that the conditional pdf exists and  that
 $f_{\theta|\xvec}(\varphi|\xvec)>0$ for almost all  $\varphi \in{\mathbb{R}}$ and $\xvec\in\chi$. For any estimator, $\hat{\theta}(\xvec):\chi\rightarrow{\mathbb{R}}$, the estimation error is $\hat{\theta}(\xvec)-\theta$ and
the corresponding $h$-outage error probability and MSE are given by $Pr\left(\left|\hat{\theta}-\theta\right|\geq \frac{h}{2}\right)$ and ${\rm E} \left[ {\left|\hat{\theta}-\theta\right|}^2\right]$, respectively.
The minimum  probability of  $h$-outage error
is (see e.g. \cite{KAY})
\begin{equation}
\label{Pe_Gen_MAP}
\min\sb{\hat{\theta}(\xvec)} Pr\left(\left|\hat{\theta}-\theta\right| >\frac{h}{2}\right)=1-{\rm E} \left[\max_{\hat{\theta}(\xvec)}\int\limits_{\hat{\theta}(\xvec)-\frac{h}{2}}^{\hat{\theta}(\xvec)+\frac{h}{2}} f_{\theta|\xvec}(\varphi|\xvec){\ud} \varphi
\right]
\end{equation}
which is attained  by the following  estimator
 \be
 \label{general_MAP} \hat{\theta}_{h}(\xvec)=\arg\max_{\hat{\theta}(\xvec)}\int_{\hat{\theta}(\xvec)-\frac{h}{2}}^{\hat{\theta}(\xvec)+\frac{h}{2}} f_{\theta|\xvec}(\varphi|\xvec){\ud} \varphi\;.
 \ee
The MAP estimator is obtained by  (\ref{general_MAP}) in the limit $h\rightarrow 0$. Thus, the estimator in  (\ref{general_MAP}), named $h$-MAP, is a generalized version of the MAP estimator for any threshold $h$. This estimator can be implemented by maximization of the conditional pdf, $f_{\theta|\xvec}(\cdot|\xvec)$ after convolution with an $h$-width rectangular window.
%In addition, it is well known that the MMSE estimator is the conditional expectation estimator, $\hat{\theta}_{MMSE}={\rm E}[\theta|\xvec ]$.
Calculation of the  minimum   $h$-outage error probability in (\ref{Pe_Gen_MAP}) as well as the MMSE is
usually not tractable.
Tight lower bounds on these performance measures   are useful for performance
analysis and system design.  Lower bounds on the outage error probability  can be useful also for derivation of MSE bounds, as will be demonstrated in Section \ref{MSE_bound}. In addition, the outage error probability lower bounds can be useful for
upper bounding the cdf  of the absolute error, given by
\be
F_{\left|\hat{\theta}-\theta\right|}\left( \frac{h}{2}\right)=Pr\left(\left|\hat{\theta}-\theta\right| \leq \frac{h}{2}\right)=1-Pr\left(\left|\hat{\theta}-\theta\right| >\frac{h}{2}\right)\;.
\ee

%%%%%%%%%%%%%%%%%%%%%%%%%%%%%%%%%%%%%%%%%%%%%%%%%%%%%%%%%%%%%%%%%%%%%%%%%%%%%%%%%%%%%

\section{A general class of lower bounds on  {\MakeLowercase{$h$}}-outage error probability }
\label{der}
\subsection{Derivation of the general class of bounds}
Let  $u_h(\xvec,\theta)$ denote an  indicator function
\begin{equation}
\label{u_def}
u_h(\xvec,\theta)= \left\{
\begin{array}{rl}
0 & \text{if } \left|\hat{\theta}-\theta\right| > \frac{h}{2}\\
1 & \text{if } \left|\hat{\theta}-\theta\right| \leq \frac{h}{2}
\end{array} \right.
\end{equation}
where   $\hat{\theta}(\xvec)$ is an  estimator of the random parameter $\theta$.
According to  reverse H$\ddot{\text{o}}$lder's inequality \cite{HLP}
\begin{equation}
\label{ineqC}
{\rm E}\left[ |u_h(\xvec,\theta)v_h(\xvec,\theta)|\right]\geq{\rm E}^{p}\left[ |u_h(\xvec,\theta)|^{\frac{1}{p}}\right]{\rm E}^{1-p}\left[ |v_h(\xvec,\theta)|^{\frac{1}{1-p}}\right],~~~\forall p>1
\end{equation}
for any arbitrary scalar function $v_h(\xvec,\theta)$ such that $v_h(\xvec,\theta)\neq 0$ for almost all $\theta\in\mathbb{R},~\xvec\in\chi$  and subject  to  the  existence of these expectations. Using (\ref{u_def}), one obtains
 \begin{equation}
\label{pe_defC}
{\rm E}\left[ \left|u_h(\xvec,\theta)\right|^{\frac{1}{p}}\right]=Pr\left(\left|\hat{\theta}-\theta\right| \leq \frac{h}{2}\right)=1-Pr\left(\left|\hat{\theta}-\theta\right| >\frac{h}{2}\right),~~\forall p>1
\end{equation}
\be
\label{uv_def}
{\rm E}\left[ \left|u_h(\xvec,\theta)v_h\left(\xvec,\theta\right)\right|\right]={\rm E}\left[\int_{\hat{\theta}-\frac{h}{2}}^{{\hat{\theta}+\frac{h}{2}}} f_{\theta|\xvec}(\varphi|\xvec)\left|v_h\left(\xvec,\varphi\right)\right|{\ud} \varphi\right] \;.
\ee
By substitution of  (\ref{pe_defC}) and (\ref{uv_def}) into
(\ref{ineqC})  one obtains the following lower bound on the  outage error probability:
\begin{eqnarray}
\label{Cbound1}
Pr\left(\left|\hat{\theta}-\theta\right| >\frac{h}{2}\right) \geq 1- {\rm E}^{\frac{1}{p}}\left[\int_{\hat{\theta}-\frac{h}{2}}^{{\hat{\theta}+\frac{h}{2}}} f_{\theta|\xvec}(\varphi|\xvec)\left|v_h\left(\xvec,\varphi\right)\right|{\ud} \varphi \right]{\rm E}^{\frac{p-1}{p}}\left[ \left|v_h(\xvec,\theta)\right|^{\frac{1}{1-p}}\right]\;.&&
\end{eqnarray}
In general, this bound is a function of  the estimator $\hat{\theta}$. The following theorem states the condition to obtain valid bounds which are independent of the estimator.
\begin{theorem}
\label{Th1}
Under the assumption that $f_{\theta|\xvec}(\theta|\xvec)>0$ for almost all $\theta\in\mathbb{R},~\xvec\in\chi$, 
 a necessary and sufficient condition for the lower bound
in (\ref{Cbound1}) to be a valid lower bound which is independent of the estimator $\hat{\theta}$, is that the function 
\be
\label{gdef}
g_h(\xvec,\theta)\define f_{\theta|\xvec}(\theta|\xvec)|v_h(\xvec,\theta)|
\ee
  is periodic in $\theta$ with period $h$, for a.e. $\xvec\in\chi$. 
  \end{theorem}
 {\em{Proof:}}  The proof appears in  Appendix A.
 
The periodic function $g_h(\xvec,\theta)$ is chosen such that  it is also piecewise continuous with respect to (w.r.t.) $\theta$, and has left and right-hand derivatives $\forall\theta\in[0,h]$. Thus, $g_h(\xvec,\theta)$ can be represented using Fourier series \cite{Fourier}:
\beqna
\label{VdefC}
g_h(\xvec,\theta)=\sum_{k=-\infty}^\infty a_k(\xvec,h)e^{i\frac{2 \pi k}{h}\theta},~~a.e.~ \xvec\in\chi\;.
\eeqna
Using (\ref{pe_defC}), (\ref{uv_def}), (\ref{gdef}), and (\ref{VdefC}) we obtain
\beqna
\label{nume}
{\rm E}\left[ |v_h(\xvec,\theta)|^{\frac{1}{1-p}}\right]={\rm E}\left[\int_{ {\mathbb{R}}} \left(\sum_{k=-\infty}^\infty a_k(\xvec,h)e^{i\frac{2 \pi k}{h} \varphi}\right)^{\frac{1}{1-p}} f_{\theta|\xvec}^{\frac{p}{p-1}}( \varphi|\xvec){\ud} \varphi\right]
\eeqna
and
\beqna
\label{denume}
{\rm E}\left[ \left|u_h(\xvec,\theta)v_h\left(\xvec,\theta\right)\right|\right]={\rm E}\left[\int_{\hat{\theta}-\frac{h}{2}}^{{\hat{\theta}+\frac{h}{2}}} g_h(\xvec,\varphi){\ud}  \varphi\right]=
{\rm E}\left[\int_{\hat{\theta}-\frac{h}{2}}^{{\hat{\theta}+\frac{h}{2}}} \sum_{k=-\infty}^\infty a_k(\xvec,h)e^{i\frac{2 \pi k}{h} \varphi}{\ud}  \varphi\right] =h{\rm E}\left[    a_0(\xvec,h)  \right]\;.
\eeqna
By substituting (\ref{denume}) and (\ref{nume}), the bound in (\ref{Cbound1}) can be rewritten as
\beqna
\label{lowerBBC}
Pr\left(\left|\hat{\theta}-\theta\right| >\frac{h}{2}\right) \geq B_{\frac{h}{2},p}
\end{eqnarray}
where
\begin{eqnarray}
\label{B_def}
 B_{\frac{h}{2},p}=1- h^{\frac{1}{p}}{\rm E}^{\frac{1}{p}}\left[   a_0(\xvec,h)\right]{\rm E}^{\frac{p-1}{p}}\left[\int\limits_{{\mathbb{R}}}\left(\sum_{k=-\infty}^\infty a_k(\xvec,h)e^{i\frac{2 \pi k}{h} \varphi}\right)^{\frac{1}{1-p}}f_{\theta|\xvec}^{\frac{p}{p-1}}( \varphi|\xvec){\ud}  \varphi\right],~~~p>1\;,
\end{eqnarray}
and
\be
\label{a0}
 a_0(\xvec,h)=\frac{1}{h}\sum_{k=-\infty}^\infty\int_0^h  a_k(\xvec,h)e^{i\frac{2 \pi k}{h} \varphi}{\ud}  \varphi=
 \frac{1}{h}\int_0^h  \sum_{k=-\infty}^\infty a_k(\xvec,h)e^{i\frac{2 \pi k}{h} \varphi}{\ud}  \varphi\;.
\ee
Using different series of functions  $\{a_k(\xvec,h)\}$ and  $p$, one obtains different bounds from this class.   It should be noted that  according to (\ref{VdefC}), $\{a_k(\xvec,h)\}$ should be chosen such that
$g_h(\xvec,\theta)$ is a positive function and in particular, $a_0(\xvec,h)>0$. In addition,  the series $\{a_k(\xvec,h)\}$ should be in $l_2({\mathbb{R}})$ where $l_2({\mathbb{R}})$ is the Hilbert space of square-summable real sequences, such that 
the Fourier series  converges.

 Since $Pr\left(\left|\hat{\theta}-\theta\right| >\frac{h}{2}\right)$ is a non-negative  non-increasing function of $h$,   the class of bounds in (\ref{B_def})  can be improved using the ``valley-filling" operator \cite{BT74}, \cite{ExtendedZZ}.
 The ``valley-filling" operator, $V$, returns
a non-increasing function  by filling
in any valleys in the input function, i.e. for function $f(\cdot)$ the operator results
\begin{equation}
\label{Vdef}
Vf(h)=\max\sb{\xi\geq 0}f(h+\xi),~~~h\geq 0\;.
\end{equation}
In addition, 
using the non-negativity property of the probability of outage error, negative values of the bound are limited to zero.

\subsection{Example - single coefficient bounds}
\label{singleC}
  In this section, an example for derivation of bounds from the proposed class by choosing specific Fourier coefficients is given.
 By substituting the choice $a_0(\xvec,h)>0$ and $a_k(\xvec,h)=0,~~~\forall k\neq 0$ in (\ref{B_def}), one obtains the class of single coefficient bounds
 \begin{eqnarray}
\label{example1}
 Pr\left(\left|\hat{\theta}-\theta\right| >\frac{h}{2}\right) \geq B_{\frac{h}{2},p}^{(1)}=1- h^{\frac{1}{p}}{\rm E}^{\frac{1}{p}}\left[   a_0(\xvec,h)\right]{\rm E}^{\frac{p-1}{p}}\left[a_0^{\frac{1}{1-p}}(\xvec,h)\int_{{\mathbb{R}}} f_{\theta|\xvec}^{\frac{p}{p-1}}( \varphi|\xvec){\ud}  \varphi\right],~~~p>1\;.
\end{eqnarray}
Now, we maximize the bound  w.r.t.  $a_0(\xvec,h)$. According to   H$\ddot{\text{o}}$lder's inequality \cite{HLP}
\be
{\rm E}^{\frac{1}{p}}\left[   a_0(\xvec,h)\right]{\rm E}^{\frac{p-1}{p}}\left[a_0^{\frac{1}{1-p}}(\xvec,h)\int_{{\mathbb{R}}} f_{\theta|\xvec}^{\frac{p}{p-1}}( \varphi|\xvec){\ud}  \varphi\right]
\geq {\rm E}\left[\left(\int_{{\mathbb{R}}} f_{\theta|\xvec}^{\frac{p}{p-1}}( \varphi|\xvec){\ud}  \varphi\right)^{\frac{p-1}{p}}\right],~~~p>1,~a_0(\xvec,h)>0
\ee
for all non-negative functions $ a_0(\xvec,h)$ and it becomes an equality {\em{iff}}
\be
\label{a0_SC}
a_0(\xvec,h)=c\left( \int_{{\mathbb{R}}} f_{\theta|\xvec}^{\frac{p}{p-1}}( \varphi|\xvec){\ud}  \varphi \right)^{\frac{p-1}{p}},
\ee
where $c$ is a positive constant. Thus, by substituting (\ref{a0_SC}) into (\ref{example1}),  one obtains the following tightest single coefficient
 bound:
 \begin{eqnarray}
 \label{sin_coef}
Pr\left(\left|\hat{\theta}-\theta\right| >\frac{h}{2}\right) \geq B_{\frac{h}{2},p}^{(1)}= 1- h^{\frac{1}{p}}{\rm E}\left[   \left( \int_{{\mathbb{R}}} f_{\theta|\xvec}^{\frac{p}{p-1}}( \varphi|\xvec){\ud}  \varphi \right)^{\frac{p-1}{p}}\right],~~~p>1\;.
\end{eqnarray} 
Since the probability of error should be non-negative, the bound in (\ref{sin_coef}) can be modified to   
 \begin{eqnarray}
 \label{sub_opt_final}
Pr\left(\left|\hat{\theta}-\theta\right| >\frac{h}{2}\right) \geq \max\left\{0,B_{\frac{h}{2},p}^{(1)}\right\}= \max\left\{0,1- h^{\frac{1}{p}}{\rm E}\left[   \left( \int_{{\mathbb{R}}} f_{\theta|\xvec}^{\frac{p}{p-1}}( \varphi|\xvec){\ud}  \varphi \right)^{\frac{p-1}{p}}\right]\right\},~~~p>1\;.
\end{eqnarray}
 %%%%%%%%%%%%%%%%%%%%%%%%%%%%%%%%%%%%%%%%%%%%%%%%%%%%%%%%%%%%%%%%%%%%%%%%%%%%%%%%%%%%%%%%%%%%%%%%%%%%%%%%%%%%
%%%%%%%%%%%%%%%%%%%%%%%%%%%%%%%%%%%%%%%%%%%%%%%%%%%%%%%%%%%%%%%%%%%%%%%%%%%%%%%%%%%%%%%%%%%%%%%%%%%%%%%%%%

%%%%%%%%%%%%%%%%%%%%%%%%%%%%%%%%%%%%%%%%%%%%%%%%%%%%%%%%%%%%%%%%%%%%%%%%%%%%%%%%%%%%%%%%%%%%%%%%%%%%%%%%%5
\section{The tightest subclass of lower bounds on the probability of outage error}
\label{tight_der}
According to   H$\ddot{\text{o}}$lder's inequality \cite{HLP}
\begin{eqnarray}
\label{try1}
{\rm E}^{\frac{1}{p}}\left[   a_0(\xvec,h)\right]{\rm E}^{\frac{p-1}{p}}\left[\int\limits_{{\mathbb{R}}}\left(\sum_{k=-\infty}^\infty a_k(\xvec,h)e^{i\frac{2 \pi k}{h} \varphi}\right)^{\frac{1}{1-p}}f_{\theta|\xvec}^{\frac{p}{p-1}}( \varphi|\xvec){\ud}  \varphi\right]\hspace{5cm}\nonumber\\\geq
 {\rm E}\left[ a_0^{\frac{1}{p}}(\xvec,h)\left(\int\limits_{{\mathbb{R}}}\left(\sum_{k=-\infty}^\infty a_k(\xvec,h)e^{i\frac{2 \pi k}{h} \varphi}\right)^{\frac{1}{1-p}}f_{\theta|\xvec}^{\frac{p}{p-1}}( \varphi|\xvec){\ud}  \varphi\right)^{\frac{p-1}{p}}\right],~~~\forall p>1
\end{eqnarray}
which becomes an equality {\em{iff}}
\beqna
\label{new_a0}
    a_0(\xvec,h)=c(h)\int\limits_{{\mathbb{R}}}\left(\sum_{k=-\infty}^\infty a_k(\xvec,h)e^{i\frac{2 \pi k}{h} \varphi}\right)^{\frac{1}{1-p}}f_{\theta|\xvec}^{\frac{p}{p-1}}( \varphi|\xvec){\ud}  \varphi
\eeqna
where  $c(h)$ denotes a positive constant independent of $\xvec$ and $\theta$. Thus, for given coefficients $\left\{a_k(\xvec,h)\right\},~k\neq 0$ the tightest subclasses of bounds in the proposed class is 
\begin{eqnarray}
\label{tightest_given}
  \tilde{B}_{\frac{h}{2},p}=1- h^{\frac{1}{p}}{\rm E}\left[ a_0^{\frac{1}{p}}(\xvec,h)\left(\int\limits_{{\mathbb{R}}}\left(\sum_{k=-\infty}^\infty a_k(\xvec,h)e^{i\frac{2 \pi k}{h} \varphi}\right)^{\frac{1}{1-p}}f_{\theta|\xvec}^{\frac{p}{p-1}}( \varphi|\xvec){\ud}  \varphi\right)^{\frac{p-1}{p}}\right]\;.
\end{eqnarray}

Let assume that the Fourier series of the functions $g_h(\xvec,\theta)$ and $g_h'(\xvec,\theta)\define\frac{\partial g_h(\xvec,\theta)}{\partial\theta}$ w.r.t. $\theta$ converges uniformly for almost all $\theta\in\mathbb{R},~\xvec\in\chi$. Under this assumption, the bound in (\ref{tightest_given})
 can be maximized w.r.t. $\left\{a_k(\xvec,h)\right\},~k\neq 0$ by equating its corresponding derivatives to zero. Under the assumption that the integration and  derivatives can be reordered, one obtains
   \beqna
\label{why}
\int_{ {\mathbb{R}}}\left(\sum_{k=-\infty}^\infty a_k^{(o)}(\xvec,h)e^{i\frac{2 \pi k}{h}\varphi}\right)^{\frac{p}{1-p}}f_{\theta|\xvec}^{\frac{p}{p-1}}( \varphi|\xvec)e^{i\frac{2 \pi m}{h} \varphi} {{\ud}}  \varphi=C(\xvec,h)\delta[m],~~~\forall m\in {\mathbb{Z}},\xvec\in\chi
\eeqna
where $\mathbb{Z}$ is the set of integers,   $\delta[\cdot]$ denotes the  Kronecker delta function, and the Fourier coefficients $\left\{ a_k^{(o)}\right\}_{k\in{\mathbb{Z}}}$ are the coefficients that maximize  the bound in (\ref{tightest_given}).
 In Appendix B, it is shown that  the stationary point satisfying (\ref{why}) yields a  maximum  of  the bound in (\ref{tightest_given}).
Under the assumption that for every $\varphi$ there is $l_0(\varphi)\in{\mathbb{Z}}$ such that
\be
\label{con_cond}
f_{\theta|\xvec}( \varphi+lh|\xvec)\leq \frac{1}{l^{(1+\alpha)\frac{p-1}{p}}},~~~\forall \varphi\in{\mathbb{R}},~ |l|>|l_0(\varphi)|,~l\in{\mathbb{Z}}\;,
\ee
for arbitrary $\alpha>0$, the series $\sum_{l=-\infty}^{\infty}f_{\theta|\xvec}^{\frac{p}{p-1}}( \varphi+lh|\xvec)$ converges for given $p$  and  the integral in the l.h.s. of (\ref{why}) can be divided into an infinite sum of integrals, where each integral is over a single period, $h$, and the delta function can be replaced by its Fourier series representation on $[0,h]$. Thus,  (\ref{why}) can be rewritten as
\begin{eqnarray}
\label{partEq}
\sum_{l=-\infty}^{\infty}\int_{lh}^{(l+1)h} \left(\sum_{k=-\infty}^\infty a_k^{(o)}(\xvec,h)e^{i\frac{2 \pi k}{h}\varphi}\right)^{\frac{p}{1-p}}f_{\theta|\xvec}^{\frac{p}{p-1}}( \varphi|\xvec)e^{i\frac{2 \pi m}{h} \varphi} {\ud} \varphi=\hspace{3.7cm}\nonumber\\=
\int_{0}^h \left(\sum_{k=-\infty}^\infty a_k^{(o)}(\xvec,h)e^{i\frac{2 \pi k}{h}\varphi}\right)^{\frac{p}{1-p}}\sum_{l=-\infty}^{\infty}f_{\theta|\xvec}^{\frac{p}{p-1}}( \varphi+lh|\xvec)e^{i\frac{2 \pi m}{h} \varphi} {\ud} \varphi=
\frac{C(\xvec,h)}{h}\int_{0}^h e^{i\frac{2 \pi m}{h} \varphi} {\ud} \varphi,
\eeqna
$\forall m\in \mathbb{Z},\xvec\in\chi$.
Using the uniqueness of the Fourier series representation for continuous functions \cite{Fourier}, \[\left(\sum_{k=-\infty}^\infty a_k^{(o)}(\xvec,h)e^{i\frac{2 \pi k}{h}\varphi}\right)^{\frac{p}{1-p}}\sum_{l=-\infty}^{\infty}f_{\theta|\xvec}^{\frac{p}{p-1}}( \varphi+lh|\xvec)=\frac{C(\xvec,h)}{h}\]  for almost all $\theta\in\mathbb{R},~\xvec\in\chi$, and thus the function $g_h(\xvec,\varphi)$ from (\ref{VdefC}) that maximizes   the bound in (\ref{tightest_given})
 can be expressed as
\be
\label{tightest_c}
g_h^{(o)}(\xvec,\theta)=\sum_{k=-\infty}^\infty a_k^{(o)}(\xvec,h)e^{i\frac{2 \pi k}{h} \theta}= \tilde{C}(\xvec,h)\left( \sum_{l=-\infty}^{\infty}f_{\theta|\xvec}^{\frac{p}{p-1}}( \theta+lh|\xvec)\right)^{\frac{p-1}{p}}
\ee
where $\tilde{C}(\xvec,h)=\left(\frac{C(\xvec,h)}{h}\right)^{\frac{1-p}{p}}$.
By substituting (\ref{tightest_c}) in  (\ref{a0}), one obtains
\be
\label{a0_new}
 a_0^{(o)}(\xvec,h)=
 \frac{1}{h}\int_0^h  \sum_{k=-\infty}^\infty a_k^{(o)}(\xvec,h)e^{i\frac{2 \pi k}{h} \varphi}{\ud}  \varphi=\frac{1}{h}\int_0^h  \tilde{C}(\xvec,h)\left( \sum_{l=-\infty}^{\infty}f_{\theta|\xvec}^{\frac{p}{p-1}}( \varphi+lh|\xvec)\right)^{\frac{p-1}{p}}{\ud}  \varphi\;.
\ee
%and thus, substitution of (\ref{tightest_c}) and (\ref{a0_new}) in (\ref{try1}), results in
%\begin{eqnarray}
%\label{try_final}
%{\rm E}^{\frac{1}{p}}\left[   a_0(\xvec,h)\right]{\rm E}^{\frac{p-1}{p}}\left[\int\limits_{{\mathbb{R}}}\left(\sum_{k=-\infty}^\infty a_k(\xvec,h)e^{i\frac{2 \pi k}{h} \varphi}\right)^{\frac{1}{1-p}}f_{\theta|\xvec}^{\frac{p}{p-1}}( \varphi|\xvec){\ud}  \varphi\right]\hspace{5cm}\nonumber\\\geq
% %{\rm E}\left[ \left(\frac{1}{h}\int_0^h  \tilde{C}(\xvec,h)\left( \sum_{l=-\infty}^{\infty}f_{\theta|\xvec}^{\frac{p}{p-1}}( \varphi+lh|\xvec)\right)^{\frac{p-1}{p}}{\ud}  \varphi\right)^{\frac{1}{p}}\left(\int\limits_0^h\tilde{C}^{\frac{1}{1-p}}(\xvec,h)\left(\sum_{l=-\infty}^{\infty}f_{\theta|\xvec}^{\frac{p}{p-1}}( \varphi+lh|\xvec)\right)^{\frac{p-1}{p}}{\ud}  \varphi\right)^{\frac{p-1}{p}}\right]\nonumber\\=
% \frac{1}{h^{\frac{1}{p}}}{\rm E}\left[ \int_0^h  \left( \sum_{l=-\infty}^{\infty}f_{\theta|\xvec}^{\frac{p}{p-1}}( \varphi+lh|\xvec)\right)^{\frac{p-1}{p}}{\ud}  \varphi\right]\;.
%\end{eqnarray}
By substituting (\ref{tightest_c}) and (\ref{a0_new}) in (\ref{tightest_given}), the  tightest subclass of bounds in this class is given by
\begin{eqnarray}
\label{final_boundCC}
B_{\frac{h}{2},p}^{(o)}=1- {\rm E}\left[ \int_0^h  \left( \sum_{l=-\infty}^{\infty}f_{\theta|\xvec}^{\frac{p}{p-1}}( \varphi+lh|\xvec)\right)^{\frac{p-1}{p}} {\ud}  \varphi\right],~~~\forall p>1\;.
\end{eqnarray}

The term $\left( \sum_{l=-\infty}^{\infty}f_{\theta|\xvec}^{\frac{p}{p-1}}( \varphi+lh|\xvec)\right)^{\frac{p-1}{p}}$ is the $\frac{p}{p-1}$ norm of $\left\{f_{\theta|\xvec}( \varphi+lh|\xvec) \right\}_{l\in{\mathbb{Z}}}$, and 
if the assumption in (\ref{con_cond}) satisfies for all $p\geq 1$,  it converges for  $p\geq 1$. Since the $\frac{p}{p-1}$ norm is an increasing function of $p$ \cite{HLP}, the class of bounds in (\ref{final_boundCC})  satisfies
\be
\label{norm_ineq}
B_{\frac{h}{2},p}^{(o)}\geq B_{\frac{h}{2},r}^{(o)},~~~\forall 1<p\leq r
\ee
and
\be
B_{\frac{h}{2},p}^{(o)}\geq \lim_{p\rightarrow \infty}B_{\frac{h}{2},p}^{(o)}=0,~~~p>1\;,
\ee  
and thus, the bound is non-negative $\forall p>1$.
In particular, for $p$  that approaches  $1$ from above, i.e.  
$p\rightarrow 1^+$, the bound in (\ref{final_boundCC}) becomes
\begin{eqnarray}
\label{tightest}
B_{\frac{h}{2},1}^{(o)}= 1- {\rm E}\left[ \int_0^h  \max\sb{l\in \mathbb{Z}} \left\{ f_{\theta|\xvec}( \varphi+lh|\xvec)\right\} {\ud}  \varphi\right]
\end{eqnarray}
which is the tightest  bound on the outage error probability in the proposed class of lower bounds.

The derivations of the proposed bounds in Sections \ref{der} and \ref{tight_der} are carried out under the assumption that
 $f_{\theta|\xvec}(\varphi|\xvec)>0$ for almost all  $\varphi \in{\mathbb{R}}$ and $\xvec\in\chi$. Extension to any conditional pdf, i.e.  $f_{\theta|\xvec}(\varphi|\xvec)\geq0$ for  all  $\varphi \in{\mathbb{R}}$ and $\xvec\in\chi$,  is performed in Appendix C.

%%%%%%%%%%%%%%%%%%%%%%%%%%%%%%%%%%%%%%%%%%%%%%%%%%%%%%%%%%%%%%%%%%%%%5
\section{General classes of lower bounds on arbitrary distortion measures  and MSE}
\label{MSE_bound}
\subsection{Derivation of the proposed class of bounds}
In similar to the extension of the  ZZLB to arbitrary distortion measures \cite{Belldoc}, the proposed outage error lower bounds in (\ref{B_def}) and (\ref{final_boundCC}) can be used to derive lower bounds on any  non-decreasing (for non-negative values) and differentiable distortion measure $D(\cdot)$ with $D(0) = 0$ and derivative $\dot{D}(\cdot)$. The expectation over the distortion measure is 
\be
\label{distortion}
{\rm E}\left[D\left({\left|\hat{\theta}-\theta\right|}\right)\right]=\frac{1}{2}\int_0^\infty\dot{D}\left(\frac{h}{2}\right)Pr\left(\left|\hat{\theta}-\theta\right|>\frac{h}{2}\right) {\ud} h\;.
\ee
For non-decreasing $D(\cdot)$ where $\dot{D}(\cdot)$
is non-negative for positive arguments, (\ref{distortion}) can be lower bounded
by bounding the probability of outage error, $Pr\left(\left|\hat{\theta}-\theta\right|\geq\frac{h}{2}\right)$.
Therefore, a general class of lower bounds on the average distortion measures is obtained by substituting  $B_{\frac{h}{2},p}$ from (\ref{B_def}) in (\ref{distortion}):
\be
\label{distortionB}
{\rm E}\left[D\left({\left|\hat{\theta}-\theta\right|}\right)\right]\geq \frac{1}{2}\int_0^\infty\dot{D}\left(\frac{h}{2}\right) B_{\frac{h}{2},p}{\ud} h,~~~p>1\;.
\ee

For example, for $D\left({\left|\hat{\theta}-\theta\right|}\right)=\left|\hat{\theta}-\theta\right|^n,~~~n\in{\mathbb{N}}$ where ${\mathbb{N}}$ is  the positive integers set, (\ref{distortion}) and (\ref{distortionB}) can be rewritten as  \cite{statistics}
\be
\label{moment}
{\rm E}\left[D\left({\left|\hat{\theta}-\theta\right|}\right)\right]={\rm E}\left[ \left|\hat{\theta}-\theta\right|^n\right]=\frac{n}{2^n}\int_0^\infty Pr\left(\left|\hat{\theta}-\theta\right|\geq \frac{h}{2}\right) h^{n-1} {\ud} h,~~~n\in{\mathbb{N}}
\ee
and
\be
{\rm E}\left[ \left|\hat{\theta}-\theta\right|^n\right]\geq\frac{n}{2^n}\int_0^\infty B_{\frac{h}{2},p} h^{n-1} {\ud} h,~~~n\in{\mathbb{N}},~p>1\;,
\ee
respectively.
Thus, the proposed lower bound on outage error probability can be used to bound any moment of the absolute error in Bayesian parameter estimation.
In particular, a new class of MSE lower bounds can be obtained by
\be
\label{known_ineq}
{\rm E}\left[ \left|\hat{\theta}-\theta\right|^2\right]=\frac{1}{2}\int_0^\infty Pr\left(\left|\hat{\theta}-\theta\right|>\frac{h}{2}\right) h {\ud} h
\ee
and 
\beqna
\label{MSE_B}
{\rm E}\left[ \left|\hat{\theta}-\theta\right|^2\right]\geq C_p\define\frac{1}{2}\int\limits_0^\infty B_{\frac{h}{2},p} h {\ud} h\;.
\eeqna
In order to obtain tighter MSE bounds, one should use tighter outage error probability lower bounds. Thus, by substituting the  tightest subclass of lower bounds on the outage error probability  from (\ref{final_boundCC}) in  (\ref{MSE_B}),
one obtains ${\rm E}\left[ \left|\hat{\theta}-\theta\right|^2\right]\geq {C_p^{(o)}}$ where
\beqna
\label{final_MSE}
{C_p^{(o)}}\define\frac{1}{2}\int\limits_0^\infty B_{\frac{h}{2},p}^{(o)} h {\ud} h=  \frac{1}{2}\int\limits_0^\infty \left(1-  {\rm E}\left[ \int\limits_0^h  \left( \sum\limits_{l=-\infty}^{\infty}f_{\theta|\xvec}^{\frac{p}{p-1}}( \varphi+lh|\xvec)\right)^{\frac{p-1}{p}} {\ud}  \varphi\right]\right) h {\ud} h, 
\eeqna
which is the tightest subclass of MSE bounds for a given $p$. By substituting  the tightest outage error probability bound, $B_{\frac{h}{2},1}^{(o)}$, from (\ref{tightest}) into (\ref{final_MSE}), one obtains
\beqna
\label{MSE_tight}
 {C_{1}^{(o)}}
\define\frac{1}{2}\int\limits_0^\infty \left(1-  {\rm E}\left[\int\limits_0^h  \max\sb{l\in \mathbb{Z}}  f_{\theta|\xvec}( \varphi+lh|\xvec) {\ud}  \varphi \right] \right) h {\ud} h
\eeqna
which is the  tightest  MSE bound in this class.

%%%%%%%%%%%%%%%%%%%%%%%%%%%%%%%%%%%%%%%%%%%%%%%%%%%%%%%%%%%%%%%%%
\subsection{Example - single coefficient MSE bounds}
 In Section \ref{singleC}, the single coefficient outage error probability is described. By substituting (\ref{sub_opt_final}) in (\ref{MSE_B}), one obtains the following MSE bound
 \beqna
{\rm E}\left[ \left|\hat{\theta}-\theta\right|^2\right]\geq \frac{1}{2}\int\limits_0^{A_p^{-p}} \left(1- h^{\frac{1}{p}}A_p \right)h {\ud} h=\frac{1}{4(2p+1)A_p^{2p}}
\eeqna
 where $A_p\define {\rm E}\left[   \left( \int_{{\mathbb{R}}} f_{\theta|\xvec}^{\frac{p}{p-1}}( \varphi|\xvec){\ud}  \varphi \right)^{\frac{p-1}{p}}\right]$. Although this bound is not tight in the general case, it may be more tractable than other existing  bounds.
%%%%%%%%%%%%%%%%%%%%%%%%%%%%%%%%%%%%%%%%%%%%%%%%%%%%%%%%%%%%%%%%%%%%%%%%%%%%%%%%%%%%%%%%%5
\section{Properties of the bounds}
\label{proper}
\subsection{Relation to the ZZLB}
\begin{theorem}
 \label{th32}
 The proposed  MSE  bound in (\ref{MSE_tight}) is always tighter than the ZZLB \cite{ExtendedZZ} for $f_{\theta|\xvec}(\theta|\xvec)>0,~~~\forall \theta\in{\mathbb{R}}$ and for almost every  $\xvec\in\chi$.
 \end{theorem}
 {\em{Proof:}} Appendix D.
 \subsection{Tightness}
\begin{theorem}
\label{th31}
 If the conditional pdf $f_{\theta|\xvec}(\cdot|\xvec)$ is
unimodal, then  the  outage error probability bound in (\ref{tightest}) coincides with  the minimum   outage error probability in  (\ref{Pe_Gen_MAP})  for every $h>0$. 
\end{theorem}
{\em{Proof:}} Appendix E.

In addition,  it was shown by Bell \cite{Belldoc} that the ZZLB in (\ref{ZZLB_A11}) coincides with the minimum MSE  when the conditional pdf  is
symmetric and unimodal.  According  to Theorem \ref{th32},
the proposed  MSE  bound in (\ref{MSE_tight}) is always tighter (or equal) than the  ZZLB. Thus, if the conditional pdf  is
symmetric and unimodal, the MSE bound in (\ref{MSE_tight}) coincides with  the minimum MSE.
 %%%%%%%%%%%%%%%%%%%%%%%%%%%%%%%%%%%%%%%%%%%%%%%%%%%%%%%%%%%%%%%%%%%%
\subsection{Dependence on the likelihood ratio function}
Let $L_{\xvec,\theta}(\xvec_0,\varphi+lh,\varphi)=\frac{f_{\xvec,\theta}( \xvec_0,\varphi+lh)}{f_{\xvec,\theta}( \xvec_0,\varphi)}$ be the joint likelihood-ratio (LR) function of the  pdf of $\xvec$ and $\theta$, $f_{\xvec,\theta}( \cdot,\cdot)$, between the points $(\xvec,\theta)=(\xvec_0,\varphi+lh)$ and $(\xvec,\theta)=(\xvec_0,\varphi)$.
Accordingly, the bounds on the probability of outage error in (\ref{final_boundCC}) and (\ref{tightest})   can be rewritten  as
\begin{eqnarray}
\label{final_boundCC_LRT}
B_{\frac{h}{2},p}^{(o)}&=&1- \int_\chi \int_0^h  \left( \sum_{l=-\infty}^{\infty}f_{\theta|\xvec}^{\frac{p}{p-1}}( \varphi+lh|\xvec=\xvec_0)\right)^{\frac{p-1}{p}} f_\xvec(\xvec_0){\ud}  \varphi {\ud}{\xvec_0}\nonumber\\%&=&1- \int_\chi \int_0^h  \left( \sum_{l=-\infty}^{\infty}f_{\xvec,\theta}^{\frac{p}{p-1}}( {\xvec_0},\varphi+lh)\right)^{\frac{p-1}{p}} {\ud}  \varphi {\ud}{\xvec_0}\nonumber\\
&=&
1- \int_\chi \int_0^h  \left( \sum_{l=-\infty}^{\infty}L_{\xvec,\theta}^{\frac{p}{p-1}}(\xvec_0,\varphi+lh,\varphi)\right)^{\frac{p-1}{p}}f_{\theta|\xvec}( \varphi|\xvec=\xvec_0)f_{\xvec}( \xvec_0) {\ud}  \varphi
{\ud} \xvec_0
\nonumber\\&=&
1- {\rm{E}}\left[ \int_0^h  \left( \sum_{l=-\infty}^{\infty}L_{\xvec,\theta}^{\frac{p}{p-1}}(\xvec,\varphi+lh,\varphi)\right)^{\frac{p-1}{p}}f_{\theta|\xvec}( \varphi|\xvec) {\ud}  \varphi\right]
\end{eqnarray}
and
\begin{eqnarray}
\label{tightest_LRT}
B_{\frac{h}{2},1}^{(o)}&=& 1- \int_\chi \int_0^h  \max\sb{l\in \mathbb{Z}} \left\{ f_{\theta|\xvec}( \varphi+lh|\xvec=\xvec_0)\right\}f_\xvec(\xvec_0) {\ud}  \varphi {\ud} \xvec_0 \nonumber\\
%&=& 1- \int_\chi \int_0^h  \max\sb{l\in \mathbb{Z}} \left\{ f_{\xvec,\theta}( \xvec_0,\varphi+lh)\right\} {\ud}  \varphi {\ud} \xvec_0 \nonumber\\
&=& 1- \int_\chi\int_0^h  \max\sb{l\in \mathbb{Z}} \left\{ L_{\xvec,\theta}(\xvec_0,\varphi+lh,\varphi)\right\} f_{\theta|\xvec}(\varphi|\xvec=\xvec_0) f_{\xvec}( \xvec_0){\ud}  \varphi{\ud}\xvec_0
\nonumber\\
&=& 1- {\rm{E}}\left[\int_0^h  \max\sb{l\in \mathbb{Z}} \left\{ L_{\xvec,\theta}(\xvec,\varphi+lh,\varphi)\right\} f_{\theta|\xvec}(\varphi|\xvec){\ud}  \varphi\right]
\end{eqnarray}
respectively, where  for the sake of simplicity we assumed that the  observation vector 
 $\xvec$ is a continuous random vector. Extension to any kind of random variable is straightforward. 
 In similar,  the MSE bounds in (\ref{final_MSE}) and (\ref{MSE_tight}) can be rewritten as 
\beqna
{C_p^{(o)}}=\frac{1}{2}\int\limits_0^\infty B_{\frac{h}{2},p}^{(o)} h {\ud} h=  \frac{1}{2}\int\limits_0^\infty \left(1- {\rm{E}}\left[ \int_0^h  \left( \sum_{l=-\infty}^{\infty}L_{\xvec,\theta}^{\frac{p}{p-1}}(\xvec,\varphi+lh,\varphi)\right)^{\frac{p-1}{p}}f_{\theta|\xvec}( \varphi|\xvec) {\ud}  \varphi
\right]\right) h {\ud} h, \hspace{-1cm}
\eeqna
and
\beqna
 {C_{1}^{(o)}}
\define\frac{1}{2}\int\limits_0^\infty \left(1- {\rm{E}}\left[\int_0^h  \max\sb{l\in \mathbb{Z}} \left\{ L_{\xvec,\theta}(\xvec,\varphi+lh,\varphi)\right\} f_{\theta|\xvec}( \varphi|\xvec){\ud}  \varphi\right]\right) h {\ud} h,
\eeqna
respectively.
Thus, the proposed bounds on the outage error probability and on the MSE depend on the LR functions.

This result is consistent with previous literature. For example,
in \cite{TT} it is shown that  some well known non-Bayesian MSE lower bounds   of unbiased
estimators
are  integral transforms of the likelihood-ratio (LR) function.
In addition, it is well known that the WWLB \cite{WWLB} is a nonlinear function of the LR function and 
the Bayesian Cram$\acute{\text{e}}$r-Rao bound and Bobrovsky-Zakai bounds are special cases of this bound. 
Furthermore, the ZZLB is a function of the likelihood ratio test of binary hypothesis testing, and therefore, it is also a
function of the LR.

%%%%%%%%%%%%%%%%%%%%%%%%%%%%%%%%%%%%%%%%%%%%%%%%%%%%%%%%%%%%%%%%%%%%%%%%%%%%%%%%%%%%%%%%
 \section{Examples}
\label{simulations}
In this section,  the performance of  the proposed  lower
bounds  derived in this paper, are evaluated via  examples.
\subsection{Example 1}
In this example,  the single coefficient bound and the tightest bound from (\ref{sub_opt_final}) and (\ref{tightest}), respectively,  are derived  for the following linear Gaussian model:
\be
x=\theta+n,~~~\theta\sim {\cal{N}}\left(\mu_\theta,\sigma_\theta^2 \right),~~~n\sim {\cal{N}}\left(0,\sigma_n^2 \right)
\ee
where $\theta$ and $n$ are statistically independent and the notation ${\cal{N}}(\mu,\sigma^2)$ represents a normal density function with mean  $\mu$ and variance $\sigma^2$.
 The conditional distribution of the parameter $\theta$ given the observation vector $\xvec$  is ${\cal{N}}\left({\rm E}[\theta|x],\sigma_{\theta|x}^2 \right)$
where $\sigma_{\theta|x}^2=\frac{\sigma_\theta^2\sigma_n^2}{\sigma_\theta^2+\sigma_n^2}$.
It can be seen that in this case the conditional pdf is symmetric and unimodal and thus,  the $h$-MAP estimator is identical to the MMSE estimator for any $h>0$. Using (\ref{Pe_Gen_MAP}), the minimum $h$-outage error probability  attained by the $h$-MAP (or MMSE) estimator, is
 \beqna
 \label{linear_gauss}
\min\sb{\hat{\theta}(\xvec)} Pr\left(\left|\hat{\theta}-\theta\right| >\frac{h}{2}\right)=1-{\rm E} \left[\int\limits_{{\rm E}[\theta|x]-\frac{h}{2}}^{{\rm E}[\theta|x]+\frac{h}{2}} \frac{1}{\sqrt{2\pi \sigma_{\theta|x}^2}}e^{-\frac{\left(\varphi-{\rm E}[\theta|x]\right)^2}{2\sigma_{\theta|x}^2}}{\ud} \varphi
\right]=1-{\rm erf}\left(\frac{h}{2\sqrt{2\sigma_{\theta|x}^2} }\right)
 \eeqna
 where $\operatorname{erf}(x) = \frac{2}{\sqrt{\pi}}\int_0^x e^{-t^2} dt$ is the error function.
The single coefficient bound   (\ref{sub_opt_final}) with $p\rightarrow 1^+$  is
 \begin{eqnarray}
 \label{single_b}
Pr\left(\left|\hat{\theta}-\theta\right| >\frac{h}{2}\right) \geq \max\left\{0,B_{\frac{h}{2},p}^{(1)}\right\}= \hspace{10cm}\nonumber\\\max\left\{0,1- h^{\frac{1}{p}}{\rm E}\left[   \left( \int_{{\mathbb{R}}} f_{\theta|\xvec}^{\frac{p}{p-1}}( \varphi|\xvec){\ud}  \varphi \right)^{\frac{p-1}{p}}\right]\right\} 
=\max\left\{0,1- h^{\frac{1}{p}}\frac{1}{\left(2 \pi \sigma_{\theta|x}^2\right)^{\frac{1}{2p}}}\left( \frac{p-1}{p}\right)^{\frac{p-1}{2p}} \right\}
,~~~p>1\;.
\end{eqnarray}
It can be seen that the single coefficient bound  with  $p\rightarrow1+$ is tight for $h<1$. However,   we obtain  tighter single coefficient bounds,  for example, using $p=1.5$ and $p=5$ for $1.5<h<3$ and $h>3$, respectively.
According to Theorem \ref{th31}, for the symmetric and unimodal  conditional pdf in this case, the tightest  bound in the proposed class presented in (\ref{tightest})  is identical to (\ref{linear_gauss}) and it is tighter than the single coefficient bound for all $h$ and $p$.
%The minimum $h$-outage error probability, the tightest proposed bound, and single coefficient bounds with $p=5,1.5$ and $p\rightarrow 1+$ are presented in  Fig. \ref{LG} as a function of $h$. 

%\begin{figure}[htb]
%\centerline{\psfig{figure=linear_gauss.eps,width=13cm}}
%\vspace{-0.5cm} \caption {
%Probability of error versus $h$  for linear Gaussian model.}\label{LG}
%\end{figure}
%%%%%%%%%%%%%%%%%%%%%%%%%%%%%%%%%%%%%%%%%%%%%%%%%%%%%%%%%%%%%%%%%%%%%%%%%%5
\subsection{Example 2}
Consider the following parameter estimation
problem with  posterior pdf:
\beqna
\label{cond_exp}
f_{\theta|x}(\varphi|x)=\frac{x}{2\lambda_1}e^{-\frac{\varphi x}{\lambda_1}}u(\varphi)+\frac{x}{2\lambda_2}e^{\frac{\varphi x}{\lambda_2}}(1-u(\varphi))
\eeqna
where $u(\cdot)$ denotes the unit step function
 and $x$ is an arbitrary positive random variable. It can be seen that this conditional pdf is  unimodal but   it is   symmetric only if  $\lambda_1= \lambda_2$.

 Table I presents the MAP, MMSE, and $\tilde{h}$-MAP estimators for this problem derived under the assumption that $\lambda_1<\lambda_2$ and the corresponding probability of $h$-outage error.
 \begin{table}[h!b!p!]
 \begin{center}
 \caption{The estimators and probabilities of error for Example 2}
\begin{tabular}{|c|c|}
	\hline
Estimator &  Probability of $h$-outage error     \\
	\hline
 $\hat{\theta}_{MAP}=0$& $\frac{1}{2}{\rm E}\left[e^{\frac{-hx}{2\lambda_1}}+e^{\frac{-hx}{2\lambda_2}}\right]$ %&$\frac{5}{8}\left( \lambda_1^2+\lambda_2^2\right)$
	\\ \hline
 $\hat{\theta}_{MMSE}=\frac{\lambda_1-\lambda_2}{2x}$& $1-\frac{1}{2}{\rm E}\left[\left(-e^{\frac{\min(a_x+{h},0)}{2\lambda_2}}+e^{\frac{-\max(a_x-{h},0)}{2\lambda_1}} -e^{\frac{\min(a_x-{h},0)}{2\lambda_2}}-e^{\frac{-\max(a_x+{h},0)}{2\lambda_1}}\right)\right]$ %& $\frac{15 \lambda_1^2 +15\lambda_2^2 +10 \lambda_1 \lambda_2}{32}$ 
	\\\hline
	 $\hat{\theta}_{\tilde{h}{\text{-}}MAP}=d_x(\tilde{h})-\frac{\tilde{h}}{2}$ & $\frac{1}{2}{\rm E}\left[ e^{\min\left(0,d_x(h)-\frac{h+\tilde{h}}{2}\right)\frac{x}{\lambda_2}}+e^{-{\max\left(0,d_x(h)+\frac{h-\tilde{h}}{2}\right)\frac{x}{\lambda_1}}}\right]$ %& 
		\\ \hline
\end{tabular}
	\label{table_1}
		\end{center}
\end{table}
The notation in this table:    $a_x=\frac{\lambda_1-\lambda_2}{x}$, $c_x(h)=\left(\log\frac{\lambda_2}{\lambda_1}+\frac{x h}{\lambda_2}\right)\frac{\lambda_1 \lambda_2}{x(\lambda_1 +\lambda_2)}$, and  \[
d_x(h)=\left\{
\begin{array}{ll} 0&\text{ if }c_x(h)<0\\
c_x(h)&\text{ if }0\leq c_x(h)\leq h\\
h&\text{ if }c_x(h)>h
\end{array}\right.\;.
\]
 Using (\ref{tightest}), the  proposed bound on the probability of $h$-outage error is
\beqna
\label{exampleB}
B_{\frac{h}{2},1}^{(o)}=\frac{1}{2}{\rm E}\left[ e^{\frac{(d_x(h)-h)x}{\lambda_2}}+e^{-\frac{d_x(h)x}{\lambda_1}}\right]\;.
\eeqna
The conditional pdf $f_{\theta|\xvec}(\cdot|\xvec)$ in (\ref{cond_exp}) is
unimodal and thus
according to Theorem \ref{th31} it is identical  to the minimum probability of error for each $h$ attained by the $\tilde{h}$-MAP estimator with $\tilde{h}=h$, presented in Table I. 
It can be seen that the $h$ outage error probabilities of the $\tilde{h}{\text{-}}MAP$ estimator with $\tilde{h}=0$ and the MAP estimator are equal.
%Thus, in this example, it can be   analytically seen that the proposed bound attains the minimum  probability of $h$-outage error  for each $h$.

%Table II presents the MSE of the above estimators, for every $\xvec$ such that the expectations in this table converge. 
% %The MSE calaculations were done
%%using   the equality  in (\ref{known_ineq}) and assuming that the integration order can be replaced.
% \begin{table}[h!b!p!]
% \begin{center}
% \caption{The estimators and the corresponding MSE for Example 2}
%\begin{tabular}{|c|c|}
%	\hline
%Estimator &  MSE    \\
%	\hline
% $\hat{\theta}_{MAP}=0$& $\left(\lambda_1^2+\lambda_2^2\right){\rm E}\left[\frac{1}{x^2}\right]$ 
%	\\ \hline
% $\hat{\theta}_{MMSE}=\frac{\lambda_1-\lambda_2}{2x}$ & $\frac{3\lambda_1^2+3\lambda_2^2+2\lambda_1\lambda_2}{4}{\rm E}\left[\frac{1}{x^2} \right]$ %& $\frac{15 \lambda_1^2 +15\lambda_2^2 +10 \lambda_1 \lambda_2}{32}$ 
%	\\\hline
%	 $\hat{\theta}_{h{\text{-}}MAP}=d_x(h)-\frac{h}{2}$ & $\left(\lambda_1^2+\lambda_2^2\right){\rm E}\left[\frac{1}{x^2}\right]-2(\lambda_1-\lambda_2) {\rm E}\left[\frac{d_x(h)-\frac{h}{2}}{2x}\right]+{\rm E}\left[\left(d_x(h)-\frac{h}{2}\right)^2\right]$%& 
%		\\ \hline
%\end{tabular}
%	\label{table_2}
%	\end{center}
%\end{table}

Consider the case of discrete distribution of 
$x$: 
\[x=\left\{{\begin{array}{rl}1~~~\text {with probability }0.5\\
2~~~\text {with probability }0.5
\end{array}}\right.\] and  $\lambda_1=1$.
The minimum $h$-outage error probability and the proposed tightest bound for this distribution of $x$  are presented in  Fig. \ref{exp1} as a function of $h$ for  distribution parameter $\lambda_2=10$. In addition, 
the outage error probabilities of the MMSE, MAP, and $h$-MAP  ($h=5$) estimators      are presented in this figure.
It can be seen that for $h\rightarrow 0$ the proposed bound approaches  the MAP outage error probability.
The MMSE estimator approaches the bound for
 $h>14$.
\begin{figure}[htb]
\centerline{\psfig{figure=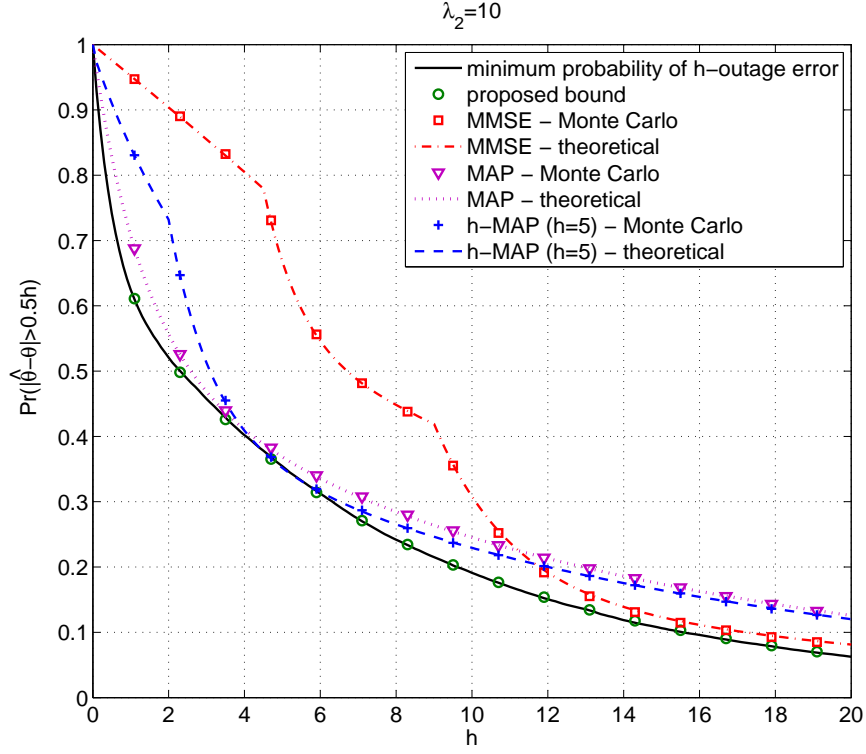,width=13cm}}
\vspace{-0.5cm} \caption {
Probability of error versus $h$ for $\lambda_2=10$ and unimodal non-symmetric exponential pdf.}\label{exp1}
\end{figure}

Fig. \ref{SNR} shows the $h$-outage error probability with $h=20$  obtained by  MMSE, MAP,  and  $h$-MAP ($h=20$) estimators compared to the proposed bound versus $\frac{1}{\lambda_2}$.
This figure shows that
 the lower bound on the $h$-outage error probability with $h=20$ coincides with
  the performance of the corresponding $h$-MAP ($h=20$) estimator.
\begin{figure}[htb]
\centerline{\psfig{figure=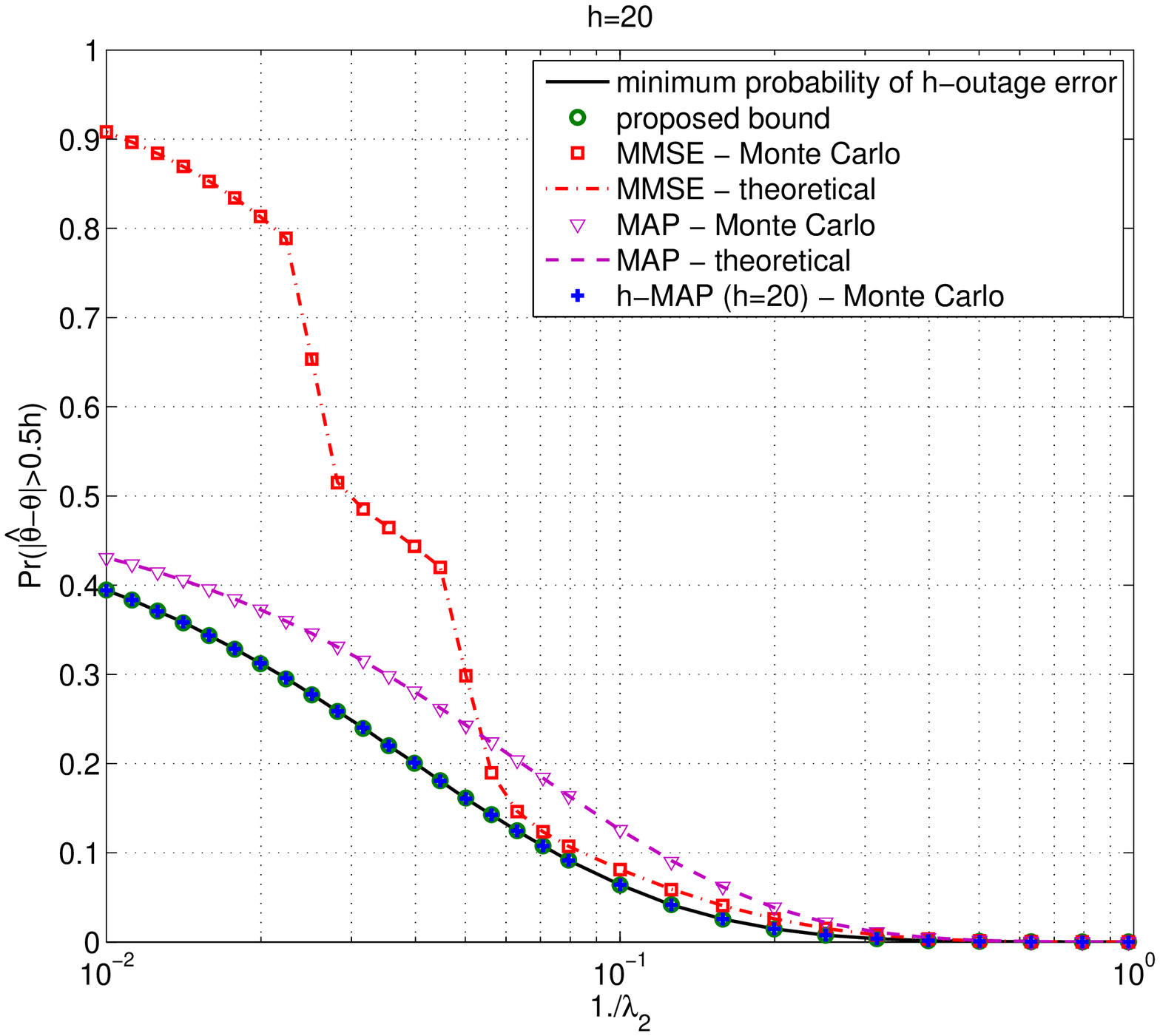,width=12.5cm}}
\vspace{-0.5cm} \caption {
Probability of $h$-outage error for $h=20$  as a function of $\frac{1}{\lambda_2}$  for unimodal non-symmetric exponential pdf.}\label{SNR}
\end{figure}

%Using  (\ref{MSE_tight}) and (\ref{exampleB}),  the analytical tightest MSE bound in the proposed class is
%\beqna
%{C_{1}^{(o)}}=\frac{5}{32}\left[\lambda_2^2+(\lambda_1\lambda_2+2\lambda_1^2)\log\frac{\lambda_2}{\lambda_1}+3\lambda_1\lambda_2+4\lambda_1^2 +\frac{1}{2}\lambda_1^2\log^2\frac{\lambda_2}{\lambda_1}\right],
%\eeqna
%for $\lambda_1<\lambda_2\;.$ %and it is coincides with the  ZZLB  \cite{ExtendedZZ} for this example.
%The MSE of the MMSE and MAP estimators and the proposed bound and the ZZLB as a function of $\frac{1}{\lambda_2}$ are presented in  Fig. \ref{RMSE}. It can be seen that for $\lambda_2$ close to $\lambda_1=1$, the proposed bound and the ZZLB become tighter. This result can be explained by the fact that  in this example the conditional pdf  becomes symmetric for $\lambda_1=\lambda_2$  and thus the MSE bound and the ZZLB  are tight at this point. 
%%
%\begin{figure}[htb]
%\centerline{\psfig{figure=MSE_Bounds.eps,width=16cm}}
%\vspace{-0.5cm} \caption {
%MSE  as function of $\frac{1}{\lambda_2}$  for unimodal non-symmetric exponential pdf.}\label{RMSE}
%\end{figure}

\subsection{Example 3}
Consider the following observation model
\be
\label{model}
x=\theta+w
\ee
where the unknown parameter, $\theta$, is distributed according to 
\[f_{\theta}(\varphi)=\frac{1}{6}\left(u(\varphi-3)-u(\varphi-6)+u(\varphi+6)-u(\varphi+3)\right)\;,\]
 $w$ is a zero-mean Gaussian random variable with  
known variance $\sigma^2$, and $\theta$ and $w$ are statistically independent.

The $h$-MAP estimator   for this problem  is   
\be
\label{h_MAP3}
\hat{\theta}_{h{\text{-}}MAP}=\left\{\begin{array}{cc}\left(6-\frac{h}{2}\right)sgn(x)& |x|>6-\frac{h}{2}\text{ or } g(h)<|x|<1.5,6<h\leq9\\
\left(3+\frac{h}{2}\right)sgn(x)&|x|<3+\frac{h}{2},h\leq 6\\
%x&3+\frac{h}{2}\leq |x|\leq6-\frac{h}{2},h<3\text{ or }|x|<g(h),|x|<1.5,6\leq h<9\text{ or }|x|<6-\frac{h}{2},9\leq h<12\\
 0& h\geq12\\
 x&\text{otherwise}\end{array}\right.\;
 \ee
 where $g(h)$ is the solution of $2erf\left(\frac{h}{2\sqrt{2\sigma^2}}\right)=erf\left(\frac{g(h)+3}{\sqrt{2\sigma^2}}\right)-erf\left(\frac{g(h)-6}{\sqrt{2\sigma^2}}\right)$ and $sgn(\cdot)$ it the sign function.
The corresponding  minimum probability of $h$-outage error is
\begin{eqnarray}
\min\sb{\hat{\theta}(\xvec)} Pr\left(\left|\hat{\theta}-\theta\right| >\frac{h}{2}\right)=\hspace{11cm}\nonumber\\\left\{ \begin{array}{lc} \frac{1}{6}\left( b(h)
-(3+h)erf\left(\frac{3+h}{\sqrt{2\sigma^2}}\right)+3erf\left(\frac{3}{\sqrt{2\sigma^2}}\right)+\sqrt{\frac{2\sigma^2}{\pi}}\left(e^{-\frac{9}{2\sigma^2}}-e^{-\frac{\left(3+h\right)^2}{2\sigma^2}}\right)
\right)& 0\leq h<3\\
\frac{1}{6}\left(3-6erf\left(\frac{6}{\sqrt{2\sigma^2}} \right)+3erf\left(\frac{3}{\sqrt{2\sigma^2}}\right)+\sqrt{\frac{2\sigma^2}{\pi}}\left(e^{-\frac{9}{2\sigma^2}}-e^{-\frac{36}{2\sigma^2}}\right)\right)&3\leq h<6\\\frac{1}{6}\left(3+3erf\left(\frac{g(h)+3}{\sqrt{2\sigma^2}}\right)-6erf\left(\frac{6-g(h)}{\sqrt{2\sigma^2}}\right)+\sqrt{\frac{2\sigma^2}{\pi}}\left(e^{-\frac{\left(g(h)+3\right)^2}{2\sigma^2}}-e^{-\frac{\left(6-g(h)\right)^2}{2\sigma^2}}\right)\right)
 &6\leq h<9\\
\frac{12-h}{6}\left(1-erf\left(\frac{h}{2\sqrt{2\sigma^2}}\right)\right)&9\leq h<12\\
0& h\geq12
\end{array}
\right.\;,
\end{eqnarray}
where $b(h)\define 6-h-(6-2h)erf\left(\frac{h}{2\sqrt{2\sigma^2}}\right)$.
It should be noticed that  the $h$-MAP estimator is not unique for $h>3$. In addition,  there is no single estimator that attains the minimum probability of $h$-outage error for every $h$.
 The MAP estimator  is obtained by  (\ref{h_MAP3}) in the limit $h\rightarrow 0$, that is
\[\hat{\theta}_{MAP}=\left\{\begin{array}{cc}6sign(x)& |x|>6\\
x&3\leq |x|\leq6\\
 3sign(x)&|x|<3\end{array}\right.\;,\] %and
%the corresponding probability of $h$-outage error for $h<6$ is $P_e(MAP)=1-\frac{1}{6}P_M$ where 
%\[P_M\define \left(\frac{h}{2}+(6-h)erf\left(\frac{h}{2\sqrt{2\sigma^2}}\right)-3erf\left(\frac{3}{\sqrt{2\sigma^2}}\right)+\left(3+\frac{h}{2}\right)erf\left(\frac{3+\frac{h}{2}}{\sqrt{2\sigma^2}}\right)+\sqrt{\frac{2\sigma^2}{\pi}}\left(e^{-\frac{\left(3+\frac{h}{2}\right)^2}{2\sigma^2}}-e^{-\frac{3^2}{2\sigma^2}}\right)\right)\;.\]
 and the MMSE estimator in this case is \[\hat{\theta}_{MMSE}=x+\frac{1}{c(x)}\sqrt{\frac{2\sigma^2}{\pi}}\left(e^{-\frac{\left(x+6\right)^2}{2\sigma^2}}-e^{-\frac{\left(x+3\right)^2}{2\sigma^2}}-e^{-\frac{\left(x-6\right)^2}{2\sigma^2}}+e^{-\frac{\left(x-3\right)^2}{2\sigma^2}}\right)\] 
 where $c(x)=erf\left(\frac{x+6}{\sqrt{2\sigma^2}}\right)-erf\left(\frac{x-6}{\sqrt{2\sigma^2}}\right)+erf\left(\frac{x-3}{\sqrt{2\sigma^2}}\right)-erf\left(\frac{x+3}{\sqrt{2\sigma^2}}\right)$.  
Using (\ref{tightest}) and the ``valley-filling" operator, the proposed bound on the outage error probability in this case is 
\begin{eqnarray}
B_{\frac{h}{2},1}^{(o)}=\left\{ \begin{array}{cc} \frac{1}{6}\left( b(h)-herf\left(\frac{\frac{h}{2}+3}{\sqrt{2\sigma^2}}\right)\right)& 0\leq h<3\\
\frac{1}{2}\left(1-erf\left(\frac{9}{2\sqrt{2\sigma^2}}\right)\right)&3\leq h<9\\
\frac{12-h}{6}\left(1-erf\left(\frac{h}{2\sqrt{2\sigma^2}}\right)\right)&9\leq h<12\\
0& h\geq12
\end{array}
\right.
\end{eqnarray}
%In Appendix D it is shown that the ZZLB can be calculated using 
%$B_{ZZLB}=\frac{1}{2}\int\limits_0^\infty   I_{ZZLB}(h)h {\ud} h$
%where \[I_{ZZLB}(h)\define V\left\{\int\limits_{-\infty}^{\infty}{\rm E}\left[\min\left(f_{\theta|\xvec}(\varphi|\xvec),f_{\theta|\xvec}(\varphi+h|\xvec)\right)\right] {\ud} \varphi\right\} \] is the Ziv-Zakai bound on the probability of $h$-outage error
%and $V$ is  the ``valley-filling" operator. 
The Ziv-Zakai probability of outage error  lower bound (defined in Appendix D) for this problem  is
\be
I_{ZZLB}(h)=\left\{ \begin{array}{cc}\frac{1}{6}\max\left(b(h)-h,3-3erf\left(\frac{9}{2\sqrt{2\sigma^2}}\right) \right)& 0\leq h<3\\
\frac{1}{2}\left(1-erf\left(\frac{9}{2\sqrt{2\sigma^2}}\right)\right)&3\leq h<9\\
\frac{12-h}{6}\left(1-erf\left(\frac{h}{2\sqrt{2\sigma^2}}\right)\right)&9\leq h<12\\
0& h\geq12\;.
\end{array}
\right.
\ee

The minimum $h$-outage error probability,  the proposed tightest bound, and the ZZLB on the probability of outage error  are presented in  Fig. \ref{last1} as a function of $h$ for   $\sigma^2=100$. In addition, 
the outage error probabilities of the MMSE, MAP, and $h$-MAP ($h=1.5$)   estimators   are presented in this figure.
It can be seen that  the proposed bound approaches  the minimum outage error probability for every $h$ and for $h<3$ there is a significant difference between the proposed bound and the outage error probability used in the ZZLB expression.
\begin{figure}[htb]
\centerline{\psfig{figure=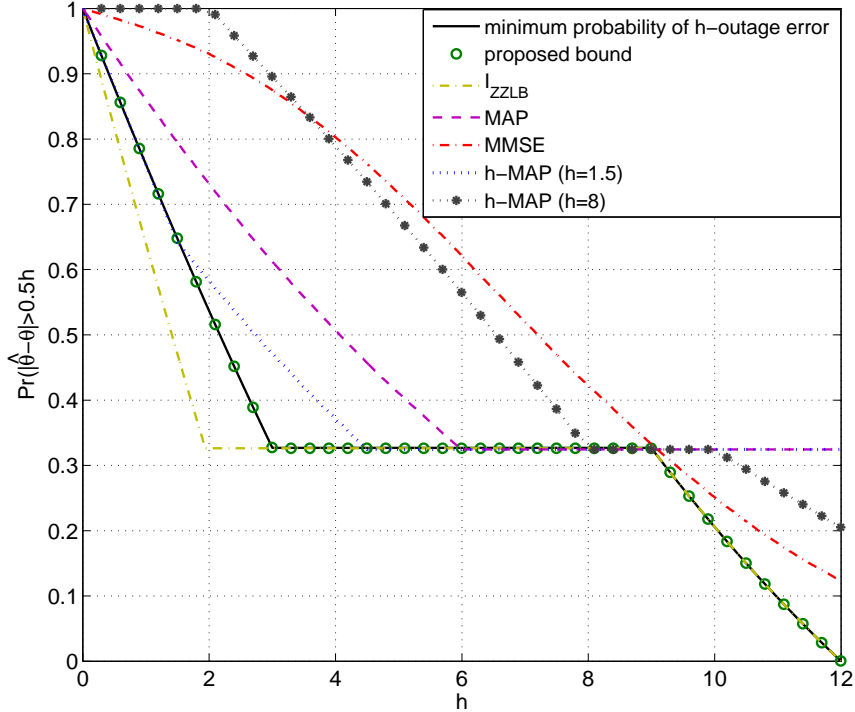,width=13cm}}
\vspace{-0.5cm} \caption {Probability of error versus $h$ for linear non-Gaussian model with $\sigma^2=100$.}\label{last1}
\end{figure}

%%%%%%%%%%%%%%%%%%%%%%%%%%%%%%%%%%%%%%%%%%%%%%%%%%%%%%%%%%%%%%%%%%%%%%%%%%%%%%%%%%%%%%%%%
\section{Conclusion}
\label{diss}
In this paper,  new classes of lower bounds on the probability of outage error and on the MSE in Bayesian parameter estimation were presented.  
The tightest subclasses of lower bounds in the proposed classes have been derived and the tightest bounds in these subclasses are presented.
For unimodal  conditional pdf the  tightest  outage error probability lower  bound     provides the minimum attainable probability of outage error.
For unimodal and symmetric conditional pdf the tightest proposed MSE bound  provides  the minimum attainable MSE.
 It is proved that the proposed MSE bound is always tighter than the well known  ZZLB. 
 The applicability of the proposed bounds was shown via examples.

%%%%%%%%%%%%%%%%%%%%%%%%%%%%%%%%%%%%%%%%%%%%%%%%%%%%%%%%%%%%%%%%%%%%%%%%%%

\section*{Appendix  A. Proof of Theorem \ref{Th1}}
\label{Appendix A}
% In this appendix, it is shown that under the assumption that $f_{\theta|\xvec}(\theta|\xvec)>0,~~~\forall \xvec\in\chi,~\varphi\in \mathbb{R}$, the expectation  ${\rm E}\left[ \left|u_h(\xvec,\theta)v_h\left(\xvec,\theta\right)\right|\right]$ is independent of the estimator $\hat{\theta}$ {\em{iff}}
% $g_h(\xvec,\theta)\define f_{\theta|\xvec}(\theta|\xvec)\cdot|v_h(\xvec,\theta)|$ is periodic in $\theta$ with period $h$, for almost  all $\xvec\in\chi$.
 
Sufficient condition: Let  $g_h(\xvec,\theta)$ be periodic in $\theta$ with period $h$, for almost every $\xvec\in\chi$.
In particular,
 \begin{equation}
\int_{\hat{\theta}-\frac{h}{2}}^{\hat{\theta}+\frac{h}{2}} g_h(\xvec,\varphi)  {\ud} \varphi=c(\xvec,h),~~~\forall \hat{\theta} 
\end{equation}
 where $c(\xvec,h)$ is not a function of the estimator $\hat{\theta}$. Thus, using (\ref{uv_def}) and (\ref{gdef})
 \begin{eqnarray}
{\rm E}\left[ \left|u_h(\xvec,\theta)v_h\left(\xvec,\theta\right)\right|\right]={\rm E}\left[\int_{\hat{\theta}-\frac{h}{2}}^{\hat{\theta}+\frac{h}{2}} g_h(\xvec,\varphi)  {\ud} \varphi \right] ={\rm E}\left[c(\xvec,h)\right]
\end{eqnarray}
 is independent of the estimator $\hat{\theta}$.

Necessary  condition:  Let  ${\rm E}\left[ \left|u_h(\xvec,\theta)v_h\left(\xvec,\theta\right)\right|\right]$ be independent of the estimator $\hat{\theta}$ and  define the following family of estimators
\be
\hat{\theta}_{A,\alpha}(\xvec)=\alpha_A {\mathbf{1}}_{\xvec \in A}
\ee
where  $\xvec$ is a  random observation vector,  the indicator function, ${\mathbf{1}}_{\xvec \in A}$, is defined as  \[{\mathbf{1}}_{\xvec \in A}=\left\{\begin{array}{cc}\alpha_A &\text{if } \xvec\in A\\
0 &\text{if } \xvec\in A^c\end{array}\right.,~~~\alpha_A\in{\mathbb{R}}\;,\]
and $A^c$ is the complementary event of $A$. Then, under the assumption that ${\rm E}\left[ \left|u_h(\xvec,\theta)v_h\left(\xvec,\theta\right)\right|\right]$ is independent of the estimator $\hat{\theta}$, in particular, for each estimator $\hat{\theta}_{A,\alpha}(\xvec)$ 
\beqna
\label{random_x}
{\rm E}\left[ \left|u_h(\xvec,\theta)v_h\left(\xvec,\theta\right)\right|\right]&=&{\rm E}\left[\int_{\hat{\theta}_{A,\alpha}(\xvec)-\frac{h}{2}}^{\hat{\theta}_{A,\alpha}(\xvec)+\frac{h}{2}} g_h(\xvec,\varphi)  {\ud} \varphi \right]\nonumber\\&=&{\rm E}\left[{\mathbf{1}}_{\xvec \in A^c}  \int_{-\frac{h}{2}}^{\frac{h}{2}}  g_h(\xvec,\varphi)  {\ud} \varphi\right]+{\rm E}\left[{\mathbf{1}}_{\xvec \in A}
\int_{\alpha_A-\frac{h}{2}}^{\alpha_A+\frac{h}{2}}g_h(\xvec,\varphi)  {\ud} \varphi \right]
\nonumber\\
&=& {\rm E}\left[\int_{-\frac{h}{2}}^{\frac{h}{2}} g_h(\xvec,\varphi)  {\ud} \varphi\right]+{\rm E}\left[{\mathbf{1}}_{\xvec \in A} \left(
\int_{\alpha_A-\frac{h}{2}}^{\alpha_A+\frac{h}{2}}g_h(\xvec,\varphi)  {\ud} \varphi -\int_{-\frac{h}{2}}^{\frac{h}{2}} g_h(\xvec,\varphi)  {\ud} \varphi\right)\right]
 \eeqna
is independent of $A$ or $\alpha_A$. Thus, 
 ${\rm E}\left[ {\mathbf{1}}_{\xvec \in A} \left(\int_{\alpha_A-\frac{h}{2}}^{\alpha_A+\frac{h}{2}}g_h(\xvec,\varphi)  {\ud} \varphi-\int_{-\frac{h}{2}}^{\frac{h}{2}}  g_h(\xvec,\varphi)  {\ud} \varphi\right)\right]$
 is identical for every $A$ and $\alpha_A$. In particular, by setting
		 $A=\emptyset$ where $\emptyset$ is the empty set, one obtains\\ ${\rm E}\left[ {\mathbf{1}}_{\xvec \in \emptyset} \left(\int_{\alpha_\emptyset-\frac{h}{2}}^{\alpha_\emptyset+\frac{h}{2}}g_h(\xvec,\varphi)  {\ud} \varphi-\int_{-\frac{h}{2}}^{\frac{h}{2}}  g_h(\xvec,\varphi)  {\ud} \varphi\right)\right]=0$, and therefore \\${\rm E}\left[ {\mathbf{1}}_{\xvec \in A} \left(\int_{\alpha_A-\frac{h}{2}}^{\alpha_A+\frac{h}{2}}g_h(\xvec,\varphi)  {\ud} \varphi-\int_{-\frac{h}{2}}^{\frac{h}{2}}  g_h(\xvec,\varphi)  {\ud} \varphi\right)\right]=0$
 for every $A$. Accordingly, 
  \be
  \label{45}
  \int_{\alpha_A-\frac{h}{2}}^{\alpha_A+\frac{h}{2}}g_h(\xvec,\varphi)  {\ud} \varphi-\int_{-\frac{h}{2}}^{\frac{h}{2}}  g_h(\xvec,\varphi)  {\ud} \varphi=0,~~~a.e. ~\xvec\in\chi,~\forall \alpha_A\in{\mathbb{R}}\;.
  \ee
 Equation (\ref{45})  indicates that under the assumption that $f_{\theta|\xvec}(\varphi|\xvec)>0,~~~\forall \xvec\in\chi,~\varphi\in \mathbb{R}$,  the term ${\rm E}\left[\int_{\hat{\theta}-\frac{h}{2}}^{\hat{\theta}+\frac{h}{2}} g_h(\xvec,\varphi)  {\ud} \varphi \right] $
 is independent of the estimator $\hat{\theta}$ only if  $g_h(\xvec,\theta)$  is a periodic function w.r.t. $\theta$ with period $h$ for a.e. $\xvec\in\chi$.
 %%%%%%%%%%%%%%%%%%%%%%%%%%%%%%%%%%%%%%%%%%%%%%%%%%%%%%%%%%%%%%%%%%%%%%%%%%%%
 \section*{Appendix B.  Eq.  (\ref{why}) yields a global maximum of the bound in (\ref{tightest_given})}
\label{Appendix B}
In this appendix,  it is shown that the stationary point of (\ref{tightest_given}) w.r.t.  the Fourier coefficients  $\left\{a_k(\xvec,h)\right\}$ is a global maximum, that is,
 it will be shown that  the Hessian matrix of (\ref{tightest_given}) w.r.t.  the coefficients $\{a_k(\xvec,h)\}_{k\in{\mathbb{Z}},k\neq 0}$  is a negative-definite matrix at the point $\{a_k^{(o)}(\xvec,h)\}_{k\in{\mathbb{Z}},k\neq 0}$ which satisfies (\ref{why}).
The derivation in this appendix is carried out under the assumption that $\xvec$ is a continuous random vector. Extension to any random vector $\xvec$  satisfying $||\xvec||_2^2<\infty$ is straightforward.   For example, if $\xvec$ is a discrete random vector with the probabilities $P(\xvec=\xvec_i)=p_i,~i=0,1,\ldots$,  the pdf $f_\xvec(\xvec_0)$ is replaced by  $p_0$ along the  proof.

Under the assumption that the integration and  derivatives can be reordered, 
the
derivatives of (\ref{tightest_given}) w.r.t. $\left\{a_m(\xvec_0,h)\right\},~~~\forall m\in {\mathbb{Z}},~ m\neq 0,~\xvec_0\in\chi$ are
\begin{eqnarray}
\label{am_der}
\frac{\partial B_{\frac{h}{2},p}}{\partial a_m(\xvec_0,h)}=\frac{1}{p}h^{\frac{1}{p}}a_0^{\frac{1}{p}}(\xvec_0,h)f_\xvec(\xvec_0)\left(\int\limits_{{\mathbb{R}}}\left(\sum_{k=-\infty}^\infty a_k(\xvec,h)e^{i\frac{2 \pi k}{h} \varphi}\right)^{\frac{1}{1-p}}f_{\theta|\xvec}^{\frac{p}{p-1}}( \varphi|\xvec){\ud}  \varphi\right)^{\frac{-1}{p}}\times\nonumber\\\int_{{\mathbb{R}}}\left(\sum_{k=-\infty}^\infty a_k(\xvec_0,h)e^{i\frac{2 \pi k}{h}\varphi}\right)^{\frac{p}{1-p}}f_{\theta|\xvec}^{\frac{p}{p-1}}( \varphi|\xvec_0)e^{i\frac{2 \pi m}{h} \varphi} {\ud}  \varphi,
\end{eqnarray}
and by equating (\ref{am_der}) to zero (note that $\int\limits_{{\mathbb{R}}}\left(\sum_{k=-\infty}^\infty a_k(\xvec,h)e^{i\frac{2 \pi k}{h} \varphi}\right)^{\frac{1}{1-p}}f_{\theta|\xvec}^{\frac{p}{p-1}}( \varphi|\xvec){\ud}  \varphi>0$ according to H$\ddot{\text{o}}$lder's inequality properties), one obtains
\begin{eqnarray}
\label{int0}
\int_{{\mathbb{R}}}\left(\sum_{k=-\infty}^\infty a_k^{(0)}(\xvec_0,h)e^{i\frac{2 \pi k}{h}\varphi}\right)^{\frac{p}{1-p}}f_{\theta|\xvec}^{\frac{p}{p-1}}( \varphi|\xvec_0)e^{i\frac{2 \pi m}{h} \varphi} {\ud}  \varphi=0,~~~\forall m\in {\mathbb{Z}},~ m\neq 0,~\xvec_0\in\chi\;.
\end{eqnarray}
%%%%%%%%%%%%%%%%%%%%%%%%%%%%%%%%%%%%%%%%%%%%%%%%%%%%%%%%%%%%%%%%%
The second order derivatives are obtained as follows
\begin{eqnarray}
\label{am_al_der}
[\Hmat]_{m,l}=\frac{\partial^2 B_{\frac{h}{2},p}}{\partial a_m(\xvec_0,h)\partial a_{-l}(\xvec_0,h)}\hspace{12cm}\nonumber\\=\frac{1}{p^2(p-1)}h^{\frac{1}{p}}a_0^{\frac{1}{p}}(\xvec_0,h)f_\xvec(\xvec_0)\left(\int\limits_{{\mathbb{R}}}\left(\sum_{k=-\infty}^\infty a_k(\xvec,h)e^{i\frac{2 \pi k}{h} \varphi}\right)^{\frac{1}{1-p}}f_{\theta|\xvec}^{\frac{p}{p-1}}( \varphi|\xvec){\ud}  \varphi\right)^{\frac{-1-p}{p}}\times\hspace{2cm}\nonumber\\\int_{{\mathbb{R}}}\left(\sum_{k=-\infty}^\infty a_k(\xvec_0,h)e^{i\frac{2 \pi k}{h}\varphi}\right)^{\frac{p}{1-p}}f_{\theta|\xvec}^{\frac{p}{p-1}}( \varphi|\xvec_0)e^{i\frac{2 \pi m}{h} \varphi} {\ud}  \varphi \int_{{\mathbb{R}}}\left(\sum_{k=-\infty}^\infty a_k(\xvec_0,h)e^{i\frac{2 \pi k}{h}\varphi}\right)^{\frac{p}{1-p}}f_{\theta|\xvec}^{\frac{p}{p-1}}( \varphi|\xvec_0)e^{-i\frac{2 \pi l}{h} \varphi} {\ud}  \varphi\nonumber\\
-\frac{1}{p-1}h^{\frac{1}{p}}a_0^{\frac{1}{p}}(\xvec_0,h)f_\xvec(\xvec_0)\left(\int\limits_{{\mathbb{R}}}\left(\sum_{k=-\infty}^\infty a_k(\xvec,h)e^{i\frac{2 \pi k}{h} \varphi}\right)^{\frac{1}{1-p}}f_{\theta|\xvec}^{\frac{p}{p-1}}( \varphi|\xvec){\ud}  \varphi\right)^{\frac{-1}{p}}\times\hspace{3cm}
\nonumber\\\int_{{\mathbb{R}}}\left(\sum_{k=-\infty}^\infty a_k(\xvec_0,h)e^{i\frac{2 \pi k}{h}\varphi}\right)^{\frac{2p-1}{1-p}}f_{\theta|\xvec}^{\frac{p}{p-1}}( \varphi|\xvec_0)e^{i\frac{2 \pi (m-l)}{h} \varphi} {\ud}  \varphi\hspace{6cm}
\end{eqnarray}
%\nonumber\\
%= c_1(\xvec_0,h)\int_{{\mathbb{R}}}c_2(\xvec_0,\varphi,h)[\Fmat]_{m,l}(\varphi) {\ud}  \varphi
for all $ m,l\in {\mathbb{Z}},~ m,l\neq 0,~\xvec_0\in\chi$, where $\Hmat$ is the Hessian matrix.
By substituting (\ref{int0}) in (\ref{am_al_der}), the Hessian matrix at  $\{a_k^{(o)}(\xvec_0,h)\}$ is
\begin{eqnarray}
\label{amal}
\left[\Hmat\left(\left\{a_k^{(o)}(\xvec_0,h)\right\}\right)\right]_{m,l}=-c_1(\xvec_0,h)\int_{{\mathbb{R}}}c_2(\xvec_0,\varphi,h)[\Fmat]_{m,l}(\varphi) {\ud}  \varphi ,
\end{eqnarray}
where
 \[c_1(\xvec_0,h)\define h^{\frac{1}{p}}a_0^{\frac{1}{p}}(\xvec_0,h)\frac{1}{p-1}f_\xvec(\xvec_0)\left(\int\limits_{{\mathbb{R}}}\left(\sum_{k=-\infty}^\infty a_k(\xvec,h)e^{i\frac{2 \pi k}{h} \varphi}\right)^{\frac{1}{1-p}}f_{\theta|\xvec}^{\frac{p}{p-1}}( \varphi|\xvec){\ud}  \varphi\right)^{\frac{-1}{p}}\;,\]  \[c_2(\xvec_0,\varphi,h)\define \left(\sum_{k=-\infty}^\infty a_k^{(o)}(\xvec_0,h)e^{i\frac{2 \pi k}{h}\varphi}\right)^{\frac{2p-1}{1-p}}f_{\theta|\xvec}^{\frac{p}{p-1}}( \varphi|\xvec_0),\] and
 $[\Fmat]_{m,l}(\varphi)=e^{i\frac{2 \pi (m-l)}{h}\varphi}$ is 
the $m,l$th Fourier coefficient  for any period  $h$ and for all $\varphi\in{\mathbb{R}}$. 
The periodic function, $\sum_{k=-\infty}^\infty a_k^{(o)}(\xvec_0,h)e^{i\frac{2 \pi k}{h}\varphi}$, is positive for almost all $\varphi \in{\mathbb{R}}$ and $\xvec_0\in\chi$ and in particular $a_0(\xvec_0,h)>0$ (according to (\ref{VdefC}) $\{a_k(\xvec,h)\}$ should be chosen such that
$\sum_{k=-\infty}^\infty a_k(\xvec_0,h)e^{i\frac{2 \pi k}{h}\varphi}$ is a positive function) and thus $c_1(\xvec_0,h)>0$ and $c_2(\xvec_0,\varphi,h)>0$
for all $h>0$, $p>0$ and for almost all $\varphi \in{\mathbb{R}}$ and $\xvec_0\in\chi$. 
For given $\varphi\in{\mathbb{R}}$, the infinite matrix of the Fourier coefficients, $\Fmat(\varphi)$,
is known to be positive-definite. Therefore, $\Hmat$ is an infinite  positive-definite matrix and (\ref{int0}) yields a maximum of (\ref{tightest_given}).
%%%%%%%%%%%%%%%%%%%%%%%%%%%%%%%%%%%%%%%%%%%%%%%%%%%%%%%%%%%%%%%%%%%%%%%%%%%%%%%%%%%55
 
\section*{Appendix C. Unknown parameter with bounded support}
The derivations in Sections \ref{der} and \ref{tight_der} are carried out under the assumption that the unknown parameter is unbounded with $f_{\theta|\xvec}(\theta|\xvec)>0,~~~\forall \theta\in {\mathbb{R}}$  for almost all $\xvec \in \chi$. In this appendix, it is shown that the derived bounds are suitable also for  parameters with  any support.  The support of the unknown parameter $\theta$ for given observation vector, $\xvec$, is defined as
\[S_{\theta|\xvec}=\overline{\left\{\varphi:f_{\theta|\xvec}(\varphi|\xvec)>0,~\varphi\in{\mathbb{R}}\right\}}\]
where $\overline{A}$ is the closure of a set $A$.

Let define the function
\be
\label{delta}
\varepsilon(\xvec,\theta)=\left\{
\begin{array}{lr}
0 & \text{if } \theta\in S_{\theta|\xvec}\\
\delta \frac{1}{\sqrt{2\pi\sigma^2}}e^{-\frac{-\theta^2}{2\sigma^2}} & \text{if }\theta\notin S_{\theta|\xvec}
\end{array} \right.
\ee
where $\delta>0$, $\sigma\gg 1$. Using this definition, it can be stated that
\be
\label{one}
{\rm E}\left[ \left|v_h(\xvec,\theta)\right|^{\frac{1}{1-p}}\right]={\rm E}\left[\int_{S_{\theta|\xvec}} f_{\theta|\xvec}(\varphi|\xvec)\left|v_h(\xvec,\varphi)\right|^{\frac{1}{1-p}} {\ud } \varphi\right]
={\rm E}\left[\int_{S_{\theta|\xvec}} \left(f_{\theta|\xvec}(\varphi|\xvec)+\varepsilon(\xvec,\varphi)\right)\left|v_h(\xvec,\varphi)\right|^{\frac{1}{1-p}} {\ud } \varphi\right]
\ee
and
\be
\label{two}
{\rm E}^{\frac{1}{p}}\left[\int_{\hat{\theta}-\frac{h}{2}}^{{\hat{\theta}+\frac{h}{2}}} f_{\theta|\xvec}(\varphi|\xvec)\left|v_h\left(\xvec,\varphi\right)\right|{\ud} \varphi \right]\leq
{\rm E}^{\frac{1}{p}}\left[\int_{\hat{\theta}-\frac{h}{2}}^{{\hat{\theta}+\frac{h}{2}}} \left(f_{\theta|\xvec}(\varphi|\xvec)+\varepsilon(\xvec,\varphi)\right)\left|v_h\left(\xvec,\varphi\right)\right|{\ud} \varphi \right],~p>1\;.
\ee
Using (\ref{Cbound1}), (\ref{one}), and (\ref{two}), 
 one obtains the following lower bound on the  outage error probability 
\begin{eqnarray}
\label{Cbound111}
Pr\left(\left|\hat{\theta}-\theta\right| >\frac{h}{2}\right)\hspace{13cm}\nonumber\\ \geq  1- {\rm E}^{\frac{1}{p}}\left[\int_{\hat{\theta}-\frac{h}{2}}^{{\hat{\theta}+\frac{h}{2}}} (f_{\theta|\xvec}(\varphi|\xvec)+\varepsilon(\xvec,\varphi))\left|v_h\left(\xvec,\varphi\right)\right|{\ud} \varphi \right]{\rm E}^{\frac{p-1}{p}}\left[ \int_{S_{\theta|\xvec}} (f_{\theta|\xvec}(\varphi|\xvec)+\varepsilon(\xvec,\varphi))\left|v_h(\xvec,\theta)\right|^{\frac{1}{1-p}}\right],
\end{eqnarray}
for all $\delta>0$, $p>1$. In similar to Theorem \ref{Th1}, define the function
\be
\tilde{g}_h(\xvec,\theta)\define \left(f_{\theta|\xvec}(\theta|\xvec)+\varepsilon(\xvec,\theta)\right)|v_h(\xvec,\theta)|\;.
\ee
A valid bound which is independent on $\hat{\theta}$ is obtained {\em{iff}} $\tilde{g}_h(\xvec,\theta)$  is periodic in $\theta$ with period $h$, for a.e. $\xvec\in\chi$ and $\theta\in {\mathbb{R}}$
and
 the periodic extension $\tilde{g}_h(\xvec,\theta)$ can be represented using Fourier series \cite{Fourier}:
\beqna
\label{VdefCC}
\tilde{g}_h(\xvec,\theta)=\sum_{k=-\infty}^\infty b_k(\xvec,h)e^{i\frac{2 \pi k}{h}\theta},~~a.e.~ \xvec\in\chi\;.
\eeqna
The derivation in Section \ref{der} is valid for bounded support where $g_h(\xvec,\theta)$ is replaced by $\tilde{g}_h(\xvec,\theta)$. The general class of lower bounds on the probability of error in the bounded support parameter estimation problem is
\begin{eqnarray}
\label{B_def_CP}
 \tilde{B}_{\frac{h}{2},p}&=&1- h^{\frac{1}{p}}{\rm E}^{\frac{1}{p}}\left[   b_0(\xvec,h)\right]{\rm E}^{\frac{p-1}{p}}\left[\int\limits_{-\infty}^{\infty}\left(\sum_{k=-\infty}^\infty b_k(\xvec,h)e^{i\frac{2 \pi k}{h} \varphi}\right)^{\frac{1}{1-p}}\left(f_{\theta|\xvec}( \varphi|\xvec)+\varepsilon(\xvec,\varphi)\right)^{\frac{p}{p-1}}{\ud}  \varphi\right],
\end{eqnarray}
for all $p>1$ and $h>0$.
The  tightest subclass of bounds in this class is given by
$Pr\left(\left|\hat{\theta}-\theta\right| >\frac{h}{2}\right) \geq \tilde{B}_{\frac{h}{2},p}^{(o)}$, $p>1,$
where
\begin{eqnarray}
\label{final_boundCC_CP}
\tilde{B}_{\frac{h}{2},p}^{(o)}=1- {\rm E}\left[ \int_0^h  \left( \sum_{l=-\infty}^{\infty}\left(f_{\theta|\xvec}( \varphi+lh|\xvec)+\varepsilon(\xvec,\varphi+lh)\right)^{\frac{p}{p-1}}\right)^{\frac{p-1}{p}} {\ud}  \varphi\right]\;.
\end{eqnarray}
Using (\ref{delta}) and taking the
limit $\delta\rightarrow 0$ in (\ref{B_def_CP}) and (\ref{final_boundCC_CP}), yields the bounds
\begin{eqnarray}
\label{B_def_CP2}
 \tilde{B}_{\frac{h}{2},p}&=&1- h^{\frac{1}{p}}{\rm E}^{\frac{1}{p}}\left[   b_0(\xvec,h)\right]{\rm E}^{\frac{p-1}{p}}\left[\int\limits_{-\infty}^{\infty}\left(\sum_{k=-\infty}^\infty b_k(\xvec,h)e^{i\frac{2 \pi k}{h} \varphi}\right)^{\frac{1}{1-p}}\left(f_{\theta|\xvec}( \varphi|\xvec)\right)^{\frac{p}{p-1}}{\ud}  \varphi\right],~~~p>1,
\end{eqnarray}
and
\begin{eqnarray}
\label{final_boundCC_CP2}
\tilde{B}_{\frac{h}{2},p}^{(o)}=1- {\rm E}\left[ \int_0^h  \left( \sum_{l=-\infty}^{\infty}f_{\theta|\xvec}^{\frac{p}{p-1}}( \varphi+lh|\xvec)\right)^{\frac{p-1}{p}} {\ud}  \varphi\right],~~~p>1\;,
\end{eqnarray}
respectively. In particular, for $p\rightarrow 1^+$, the bound in (\ref{final_boundCC_CP2}) becomes
\begin{eqnarray}
\label{tightest_CP}
\tilde{B}_{\frac{h}{2},1}^{(o)}= 1- {\rm E}\left[ \int_0^h  \max\sb{l\in \mathbb{Z}} \left\{ f_{\theta|\xvec}( \varphi+lh|\xvec)\right\} {\ud}  \varphi\right]
\end{eqnarray}
which is the tightest  bound on the outage error probability in the proposed class of lower bounds. Accordingly, the proposed bounds can be applied to any parameter estimation problem with unknown continuous random variable.

%%%%%%%%%%%%%%%%%%%%%%%%%%%%%%%%%%%%%%%%%%%%%%%%%%%%%%%%%%%%%%%%%%%%%%%%%%%%%%%%%%%%%%%%%%%%%
\section*{Appendix D. The proposed MSE bound is tighter than the ZZLB}
\label{MSE_Vs_ZZLB}
In this appendix, it is analytically shown  that the proposed  MSE  bound in (\ref{MSE_tight}) is always tighter than the ZZLB.
 For the sake of simplicity,  we will assume   that $\xvec$ is continuous random variable.  Extension to any random vector $\xvec$  satisfying $||\xvec||_2^2<\infty$ is straightforward.
The   MSE lower bound  from (\ref{MSE_tight}) can be rewritten as
\beqna
\label{tosefet}
 {C_{1}^{(o)}}= \frac{1}{2}\int\limits_0^\infty {\rm E}\left[\int\limits_{-\infty}^{\infty}f_{\theta|\xvec}( \varphi|\xvec) {\ud}  \varphi
-  \int\limits_0^h  \max\sb{l\in \mathbb{Z}}  f_{\theta|\xvec}( \varphi+lh|\xvec) {\ud}  \varphi \right]  h {\ud} h\;.
\eeqna
 Using the convergence condition in (\ref{con_cond}),  the first integral in (\ref{tosefet}) can be divided into infinite sum of integrals, where each integral is over a single period and (\ref{tosefet}) can be rewritten as
\beqna
\label{MSE_tight1}
C_{1}^{(o)}
=\frac{1}{2}\int\limits_0^\infty {\rm E}\left[\int\limits_0^h\min\sb{l\in \mathbb{Z}}\sum_{\stackrel{k=-\infty}{k\neq l}}^{\infty}f_{\theta|\xvec}( \varphi+kh|\xvec) {\ud}  \varphi \right] h {\ud} h\;.
\eeqna

The ZZLB (without the "valley-filling" function) is \cite{ExtendedZZ}
\begin{eqnarray}
\label{ZZLB_A11}
 B_{ZZLB}=\frac{1}{2}\int\limits_0^\infty \left( \int\limits_{-\infty}^{\infty}\left(f_\theta(\varphi)+f_\theta(\varphi+h)\right) P_{min}(\varphi,\varphi+h){\ud} \varphi\right) h {\ud} h
\end{eqnarray}
where  $f_\theta(\cdot)$ is the {\em a-priori} pdf of $\theta$
 and $P_{min}(\varphi,\varphi+h)$ is the minimum probability of error for the following detection problem:
\begin{equation}
\label{detection_prob}
{\begin{array}{rl}
&H_0: f_{\xvec|H_0}(\xvec)= f_{\xvec|\theta}(\xvec|\varphi)\\
&H_1: f_{\xvec|H_1}(\xvec)= f_{\xvec|\theta}(\xvec|\varphi+h)
\end{array}}
\end{equation}
 where $f_{\xvec|H_i}(\cdot),~i=1,2$ is the pdf of the observation vector $\xvec$ under each hypothesis, and the  prior probabilities are
\be
\label{priorP}
P(H_0)=\frac{f_{\theta}(\varphi)}{f_{\theta}(\varphi)+f_{\theta}(\varphi+h)},~~~P(H_1)=1-P(H_0)\;.
\ee
It is well known that the minimum  probability of error is obtained by the MAP criterion and for binary hypothesis testing it is given by \cite{feder}
\begin{equation}
\label{Pe_MAP}
P_{min}(\varphi,\varphi+h)=1-{\rm E}\left[\max\sb{i=1,2}P(H_i|\xvec)\right]={\rm E}\left[\min\sb{i=1,2}P(H_i|\xvec)\right]\;. 
\end{equation}
By substituting (\ref{detection_prob}) - (\ref{priorP}) in (\ref{Pe_MAP}),
the minimum  probability of error obtained by the MAP detector is
\beqna
\label{Pmin_ZZLB1}
P_{min}(\varphi,\varphi+h)={\rm E}\left[\min\left(\frac{f_{\xvec|\theta}(\xvec|\varphi)f_{\theta}(\varphi)}{\left(f_{\theta}(\varphi)+f_{\theta}(\varphi+h)\right)},\frac{f_{\xvec|\theta}(\xvec|\varphi+h)f_{\theta}(\varphi+h)}{f_\xvec(\xvec)\left(f_{\theta}(\varphi)+f_{\theta}(\varphi+h)\right)}\right)\right]\;.
\eeqna
Note that  since $\xvec$ is the only random variable  in  (\ref{Pmin_ZZLB1}),  the expectation is performed w.r.t. $\xvec$.
By substituting (\ref{Pmin_ZZLB1}) in (\ref{ZZLB_A11}), one obtains
\begin{eqnarray}
\label{ZZLB_A22}
 B_{ZZLB}&=&\frac{1}{2}\int\limits_0^\infty  \int\limits_{-\infty}^{\infty} {\rm E}\left[\min\left(\frac{f_{\xvec|\theta}(\xvec|\varphi)f_{\theta}(\varphi)}{f_\xvec(\xvec)},\frac{f_{\xvec|\theta}(\xvec|\varphi+h)f_{\theta}(\varphi+h)}{f_\xvec(\xvec)}\right)\right] {\ud} \varphi h {\ud} h\nonumber\\
 &=&\frac{1}{2}\int\limits_0^\infty  \int\limits_{-\infty}^{\infty} {\rm E}\left[\min\left(f_{\theta|\xvec}(\varphi|\xvec),f_{\theta|\xvec}(\varphi+h|\xvec)\right)\right] {\ud} \varphi h {\ud} h\;.
 \end{eqnarray}
Since $\int\limits_{-\infty}^{\infty} {\rm E}\left[\min\left(f_{\theta|\xvec}(\varphi|\xvec),f_{\theta|\xvec}(\varphi+h|\xvec)\right)\right] {\ud} \varphi \leq \int\limits_{-\infty}^{\infty} {\rm E}\left[f_{\theta|\xvec}(\varphi|\xvec)\right] {\ud} \varphi=1$,  the inner integral in the r.h.s. of (\ref{ZZLB_A22}) converges (i.e. the ZZLB  does not need a convergence condition on the pdf). Therefore, it is possible to change the order of integration w.r.t. $\xvec$ and $\varphi$, and the integral   can be divided into an infinite sum of integrals. Thus,
 \begin{eqnarray}
\label{ZZLB_A33}
 B_{ZZLB}&=&\frac{1}{2}\int\limits_0^\infty   {\rm E}\left[\int\limits_{-\infty}^{\infty}\min\left(f_{\theta|\xvec}(\varphi|\xvec),f_{\theta|\xvec}(\varphi+h|\xvec)\right){\ud} \varphi\right]  h {\ud} h\nonumber\\&=&\frac{1}{2}\int\limits_0^\infty {\rm E}\left[\sum_{l=-\infty}^{\infty}\int_{lh}^{(l+1)h} \min\left(f_{\theta|\xvec}(\varphi|\xvec),f_{\theta|\xvec}\left(\varphi+h|\xvec\right)\right){\ud}\varphi  \right] h {\ud} h\nonumber\\
 &=&\frac{1}{2}\int\limits_0^\infty {\rm{E}}\left[\int\limits_{0}^{h}\sum\limits_{l=-\infty}^{\infty} \min\left(f_{\theta|\xvec}(\varphi+lh|\xvec),f_{\theta|\xvec}(\varphi+(l+1)h|\xvec)\right) {\ud} \varphi\right] h {\ud} h\;.
\end{eqnarray}
Thus, the ZZLB can be written as
$B_{ZZLB}=\frac{1}{2}\int\limits_0^\infty   I_{ZZLB}(h)h {\ud} h$
where 
\be
\label{ZZLBI}
I_{ZZLB}(h)\define V\left\{\int\limits_{-\infty}^{\infty}{\rm E}\left[\min\left(f_{\theta|\xvec}(\varphi|\xvec),f_{\theta|\xvec}(\varphi+h|\xvec)\right)\right] {\ud} \varphi\right\} 
\ee
is referred in this paper as the Ziv-Zakai outage error probability lower bound.

In \cite{Belldoc} it is shown   that $
\min\limits\sb{k}\sum\limits_{n=0,n\neq k}^{M-1}a_n \geq \sum\limits_{n=0}^{M-2}\min(a_n ,a_{n+1})$
for any $M$ non-negative numbers $\{a_n\}_{n=0}^{M-1}$.
In a similar manner, it can be shown that
 for an infinite countable set of non-negative numbers $\{a_n\}_{n\in \mathbb{Z}}$ satisfying  $\lim\limits_{n \rightarrow \pm \infty} a_n=0$ and $\sum\limits_{n=-\infty}^{\infty}a_n<\infty$, \[
\min_{k\in \mathbb{Z}}\sum_{n =-\infty,~n\neq k}^{\infty}a_n \geq \sum_{n=-\infty}^{\infty}\min(a_n ,a_{n+1})\;.\]
In particular, under the convergence condition in (\ref{con_cond})
\beqna
\label{ZZLBVsUs}
\min\sb{k\in \mathbb{Z}}\sum_{l=-\infty, l\neq k}^{\infty}f_{\theta|\xvec}( \varphi+lh|\xvec)\geq \sum\limits_{l=-\infty}^{\infty} \min\left(f_{\theta|\xvec}(\varphi+lh|\xvec),f_{\theta|\xvec}(\varphi+(l+1)h|\xvec)\right)
\eeqna
and thus
\beqna
\label{ZZLBVsUs2}
{\rm E}\left[\min\sb{k\in \mathbb{Z}}\sum_{\stackrel{l=-\infty} {l\neq k}}^{\infty}f_{\theta|\xvec}( \varphi+lh|\xvec)\right]\geq {\rm E}\left[\sum\limits_{l=-\infty}^{\infty} \min\left(f_{\theta|\xvec}(\varphi+lh|\xvec),f_{\theta|\xvec}(\varphi+(l+1)h|\xvec)\right)\right]~\;.
\eeqna
Therefore, from  (\ref{MSE_tight1}), (\ref{ZZLB_A22}) and (\ref{ZZLBVsUs2}), one concludes that
\be
\label{MSE_ZZLB}
C_{1}^{(o)}\geq B_{ZZLB}\;.
\ee
Thus, the proposed lower bound in (\ref{MSE_tight}) is always tighter than the ZZLB in (\ref{ZZLB_A11}). Applying the  "valley-filling" operator on both sides of (\ref{MSE_ZZLB}) does not change this result.

%%%%%%%%%%%%%%%%%%%%%%%%%%%%%%%%%%%%%%%%%%%%%%%%%%%%%%%%%%%%%%%%%%%%%%

\section*{Appendix E. The tightness of the bound for  unimodal   pdf}
\label{unimodal_app_NS}
In this appendix  it is shown that if the conditional pdf $f_{\theta|\xvec}(\cdot|\xvec)$ is a
 unimodal function,    the  outage error probability bound in (\ref{tightest}) coincides with  the minimum probability of outage error in (\ref{Pe_Gen_MAP})  for every $h>0$. 
Assume that  $f_{\theta|\xvec}(\cdot|\xvec)$ is unimodal with  maximum point  $\hat{\theta}_{MAP}(\xvec)=\hat{\theta}_0(\xvec)$. 
The bound in (\ref{tightest})  involves the computation of 
	\beqna
	\label{proof1_C2}
	{\rm E}\left[ \int_0^h  \max\sb{l\in \mathbb{Z}} \left\{ f_{\theta|\xvec}\left( \varphi+lh|\xvec\right)\right\} {\ud}  \varphi\right]=	{\rm E}\left[ \int_0^h  \max\sb{l\in \mathbb{Z}} \left\{ f_{\theta|\xvec}\left( \hat{\theta}_0(\xvec)+h\left(l-\frac{\hat{\theta}_0(\xvec)-\varphi}{h}\right)|\xvec\right)\right\} {\ud}  \varphi\right]\;.
	\eeqna
Since the conditional pdf, $f_{\theta|\xvec}\left(\cdot|\xvec\right)$, is unimodal with maximum at $\hat{\theta}_0(\xvec)$, then
\beqna
\label{argmaxZ_NS}
\max\sb{l\in \mathbb{Z}} \left\{ f_{\theta|\xvec}\left( \hat{\theta}_0(\xvec)+h\left(l-\frac{\hat{\theta}_0(\xvec)-\varphi}{h}\right)|\xvec\right)\right\}=\hspace{7cm}\nonumber\\=
\max\left\{  f_{\theta|\xvec}\left( \hat{\theta}_0(\xvec)+ha_\xvec(\varphi)|\xvec\right), f_{\theta|\xvec}\left( \hat{\theta}_0(\xvec)+h\left(a_\xvec(\varphi)+1\right)|\xvec\right)\right\}
	 \eeqna
	 for all $\xvec$ and  $\varphi\in(0,h)$  where $a_\xvec(\varphi)=\left\lfloor \frac{\hat{\theta}_0(\xvec)-\varphi}{h}\right\rfloor-\frac{\hat{\theta}_0(\xvec)-\varphi}{h}$ and $\left\lfloor \cdot \right\rfloor$  is the floor  function. Note that the function $a_\xvec(\cdot)$ is continuous almost everywhere for $\varphi\in(0,h)$ and $\xvec\in\chi$ and in the continuous region its derivative is $\frac{{\ud} a_\xvec(\varphi)}{{\ud} \varphi}=\frac{1}{h}$. By substituting (\ref{argmaxZ_NS}) in (\ref{proof1_C2}) and changing variables to  $\varphi'=a_\xvec(\varphi)$, one obtains 
	 \beqna
	\label{proof2_C2}
{\rm E}\left[ \int_0^h  \max\sb{l\in \mathbb{Z}} \left\{ f_{\theta|\xvec}( \varphi+lh|\xvec)\right\} {\ud}  \varphi\right]=\hspace{8cm}\nonumber\\={\rm E}\left[ \int_0^h  \max\left\{  f_{\theta|\xvec}\left( \hat{\theta}_0(\xvec)+h a_\xvec(\varphi)|\xvec\right), f_{\theta|\xvec}\left( \hat{\theta}_0(\xvec)+h\left(a_\xvec(\varphi)+1\right)|\xvec\right)\right\} {\ud}  \varphi\right]\nonumber\\
={\rm E}\left[ \int_{-1}^0 h \max\left\{  f_{\theta|\xvec}\left( \hat{\theta}_0(\xvec)+h \varphi'|\xvec\right), f_{\theta|\xvec}\left( \hat{\theta}_0(\xvec)+h\varphi'+h|\xvec\right)\right\} {\ud}  \varphi'\right]\;.\hspace{1cm}
\eeqna
Changing variables to  $\varphi=h\varphi'+\hat{\theta}_0(\xvec)$ and decomposing the integral  to two regions, results  in
\beqna
\label{proof2_C33}
{\rm E}\left[ \int_0^h  \max\sb{l\in \mathbb{Z}} \left\{ f_{\theta|\xvec}( \varphi+lh|\xvec)\right\} {\ud}  \varphi\right]={\rm E} \left[\int_{\hat{\theta}_0(\xvec)-h}^{\hat{\theta}_0(\xvec)} \max \left\{f_{\theta|\xvec}(\varphi+h|\xvec),f_{\theta|\xvec}(\varphi|\xvec)\right\}{\ud} \varphi
\right]=\hspace{3.1cm}\nonumber\\={\rm E} \left[\int_{\hat{\theta}_0(\xvec)-h}^{\hat{\theta}_h(\xvec)-\frac{h}{2}} \max \left\{f_{\theta|\xvec}(\varphi'+h|\xvec),f_{\theta|\xvec}(\varphi'|\xvec)\right\}{\ud} \varphi'
\right]+{\rm E} \left[\int_{\hat{\theta}_h(\xvec)-\frac{h}{2}}^{\hat{\theta}_0(\xvec)} \max \left\{f_{\theta|\xvec}(\varphi|\xvec),f_{\theta|\xvec}(\varphi+h|\xvec)\right\}{\ud} \varphi
\right]\nonumber\\={\rm E} \left[\int_{\hat{\theta}_0(\xvec)}^{\hat{\theta}_h(\xvec)+\frac{h}{2}} \max \left\{f_{\theta|\xvec}(\varphi|\xvec),f_{\theta|\xvec}(\varphi-h|\xvec)\right\}{\ud} \varphi
\right]+{\rm E} \left[\int_{\hat{\theta}_h(\xvec)-\frac{h}{2}}^{\hat{\theta}_0(\xvec)} \max \left\{f_{\theta|\xvec}(\varphi|\xvec),f_{\theta|\xvec}(\varphi+h|\xvec)\right\}{\ud} \varphi
\right]\hspace{0.4cm}
\end{eqnarray}
where the $h$-MAP estimator, $\hat{\theta}_h(\xvec)$, is defined in (\ref{general_MAP}) which maximizes the area under the curve of the conditional pdf for  given length $h$. Thus,
in the unimodal case, the 
$h$-MAP estimator is the unique  estimator that satisfies   the equation $f_{\theta|\xvec}\left(\hat{\theta}_h(\xvec)-\frac{h}{2}|\xvec\right)=f_{\theta|\xvec}\left(\hat{\theta}_h(\xvec)+\frac{h}{2}|\xvec\right)$ for all $h>0$.
 Using the unimodal property, the conditional pdf satisfies
 \[
 \left\{\begin{array}{cl} f_{\theta|\xvec}(\varphi-h|\xvec)\leq f_{\theta|\xvec}(\varphi|\xvec)& \hat{\theta}_0(\xvec)<\varphi\leq \hat{\theta}_h(\xvec)+\frac{h}{2}\\f_{\theta|\xvec}(\varphi+h|\xvec)\leq f_{\theta|\xvec}(\varphi|\xvec)&\hat{\theta}_h(\xvec)-\frac{h}{2}\leq\varphi<\hat{\theta}_0(\xvec)
\end{array}\right.\;,
 \]
%  $f_{\theta|\xvec}(\varphi+h|\xvec)\leq f_{\theta|\xvec}(\varphi|\xvec)$ for all $\hat{\theta}_h(\xvec)-\frac{h}{2}\leq\varphi<\hat{\theta}_0(\xvec)$ and $f_{\theta|\xvec}(\varphi-h|\xvec)\leq f_{\theta|\xvec}(\varphi|\xvec)$ for all $\hat{\theta}_0(\xvec)<\varphi\leq \hat{\theta}_h(\xvec)+\frac{h}{2}$. 
for all $\xvec\in\chi$ and  (\ref{proof2_C33}) can be rewritten as 
\beqna
\label{unimodal_NS_Pe2}
{\rm E}\left[ \int_0^h  \max\sb{l\in \mathbb{Z}} \left\{ f_{\theta|\xvec}( \varphi+lh|\xvec)\right\} {\ud}  \varphi\right]&=&{\rm E} \left[\int_{\hat{\theta}_0(\xvec)}^{\hat{\theta}_h(\xvec)+\frac{h}{2}} f_{\theta|\xvec}(\varphi|\xvec){\ud} \varphi
\right]+{\rm E} \left[\int_{\hat{\theta}_h(\xvec)-\frac{h}{2}}^{\hat{\theta}_0(\xvec)} f_{\theta|\xvec}(\varphi|\xvec){\ud} \varphi
\right]\nonumber\\&=&
{\rm E} \left[\int_{\hat{\theta}_h(\xvec)-\frac{h}{2}}^{\hat{\theta}_h(\xvec)+\frac{h}{2}} f_{\theta|\xvec}(\varphi|\xvec){\ud} \varphi
\right]\;.
\eeqna
By substituting (\ref{unimodal_NS_Pe2}) in (\ref{tightest}),
the proposed bound on the $h$-outage error probability in the unimodal case is
\beqna
\label{unimodal_B}
B_{\frac{h}{2},1}^{(o)}=1-
{\rm E} \left[\int_{\hat{\theta}_h(\xvec)-\frac{h}{2}}^{\hat{\theta}_h(\xvec)+\frac{h}{2}} f_{\theta|\xvec}(\varphi|\xvec){\ud} \varphi
\right]
\eeqna
which is identical to the minimum probability of $h$-outage error
in (\ref{Pe_Gen_MAP}) obtained by the $h$-MAP estimator. 
Thus, the proposed  outage error probability bound in (\ref{tightest})   coincides with  the minimum probability of outage error in (\ref{Pe_Gen_MAP})  for every $h>0$   for unimodal conditional pdf.

\section*{Acknowledgment}
The authors would like to thank Dr. G. Cohen for helpful discussions
during this work. 
This research was partially supported by THE ISRAEL SCIENCE 
FOUNDATION (grant No. 1311/08) and by the Yaakov ben Yitzhak scholarship.

%Your paper has been submitted and your confirmation number is 23882. Within a few weeks the corresponding author will be contacted with additional information.

\bibliographystyle{IEEEtran}
\bibliography{errorbound4}

\end{document}